\documentclass{aa}  

\usepackage{graphicx}
\usepackage{txfonts}
\usepackage{verbatim}
\usepackage{natbib}
\usepackage{array}
\usepackage{threeparttable}
\def\kms{km\,s$^{-1}$}
\def\cmss{cm\,s$^{-2}$}

\def\msun{M$_{\odot}$}
\def\msol{M$_{\odot}$}
\def\msolyr{M$_{\odot}$\,yr$^{-1}$}

\def\rsun{R$_{\odot}$}

\def\l{$\lambda$}

\def\hea{He\,{\sc i}}
\def\heb{He\,{\sc ii}}

\def\fw{\textsc{fastwind} }
\def\phoebe{\textsc{phoebe} }
\def\spamms{\textsc{spamms} }

\begin{document}

%   \title{Spectroscopic Characterization of Three Known Massive Overcontact Binaries}
   \title{Constraining the Overcontact Phase in Massive Binary Evolution}

   \subtitle{I. Mixing in V382 Cyg, VFTS 352, and OGLE SMC-SC10 108086}

   \author{Michael Abdul-Masih\inst{1,2},
          Hugues Sana\inst{2},
          Calum Hawcroft\inst{2},
          Leonardo A. Almeida\inst{3,4},
          Sarah A. Brands\inst{5},
          Selma E. de Mink\inst{6,5,7},
          Stephen Justham\inst{5,8},
          Norbert Langer\inst{9},
          Laurent Mahy\inst{2,10},
          Pablo Marchant\inst{2},
          Athira Menon\inst{5},
          Joachim Puls\inst{11},
          \and
          Jon Sundqvist\inst{2}
          %\fnmsep\thanks{Just to show the usage of the elements in the author field}
          }
   
   \institute{
            European Southern Observatory, Alonso de Cordova 3107, Vitacura, Casilla 19001, Santiago de Chile, Chile\\
            \email{michael.abdul-masih@eso.org}
        \and
            Institute of Astrophysics, KULeuven, Celestijnenlaan 200 D, 3001 Leuven, Belgium
        \and
            Escola de Ci\^encias e Tecnologia, Universidade Federal do Rio Grande do Norte, Natal, RN 59072-970, Brazil 
        \and
            Departamento de F\'isica, Universidade do Estado do Rio Grande do Norte, Mossor\'o, RN 59610-210, Brazil
        \and
            Astronomical Institute Anton Pannekoek, Amsterdam University, Science Park 904, 1098~XH, Amsterdam, The Netherlands
        \and
            Max-Planck-Institut für Astrophysik, Karl-Schwarzschild-Straße 1, 85740 Garching bei München, Germany
        \and 
            Harvard-Smithsonian Center for Astrophysics, Harvard University, 60 Garden St, Cambridge, MA 02138, USA
        \and
            School of Astronomy \& Space Science, University of the Chinese Academy of Sciences, Beijing 100012, China
        \and
            Argelander-Institut f\"ur Astronomie, Universit\"at Bonn, Auf dem H\"ugel 71, 53121 Bonn, Germany
        \and
            Royal Observatory of Belgium, Avenue Circulaire 3, B-1180 Brussel, Belgium
        \and
            LMU München, Universitätssternwarte, Scheinerstr. 1, 81679 München, Germany
             }

%   \date{Received September 15, 1996; accepted March 16, 1997}

% \abstract{}{}{}{}{} 
% 5 {} token are mandatory
 
  \abstract
   {As potential progenitors of several exotic phenomena including gravitational wave sources, magnetic stars, and Be stars, close massive binary systems probe a crucial area of the parameter space in massive star evolution. Despite the importance of these systems, large uncertainties regarding the nature and efficiency of the internal mixing mechanisms still exist.}
   {In this work, we aim to provide robust observational constraints on the internal mixing processes by spectroscopically analyzing a sample of three massive overcontact binaries at different metallicities.}
   {Using optical phase-resolved spectroscopic data, we perform an atmosphere analysis using more traditional 1D techniques and using state-of-the-art 3D techniques.  We compare and contrast the assumptions and results of each technique and investigate how the assumptions affect the final derived atmospheric parameters. }
   %We show that \spamms is able model many different system geometries (including rotating single stars, detached binaries, semi-detached binaries, overcontact systems, etc.) and is able to reproduce observed three-dimensional phenomena.  
   {We find that in all three cases, both components of system are highly overluminous indicating either efficient internal mixing of helium or previous non-conservative mass transfer.  However, we do not find strong evidence of helium or CNO surface abundance changes usually associated with mixing.  Additionally, we find that in unequal mass systems, the measured effective temperature and luminosity of the less massive component places it very close to the more massive component on the Hertzsprung-Russell diagram.  These results were obtained independently using both of the techniques mentioned above, which suggests that these measurements are robust.}
   {The observed discrepancies between the temperature and the surface abundance measurements when compared to theoretical expectations indicate that unaccounted for additional physical mechanisms may be at play. }
   %{\spamms is a very powerful spectroscopic analysis tool that can account for the three-dimensional shapes of many massive star systems better than other currently available methods.  }

	\titlerunning{Overcontact Binaries}
	\authorrunning{M. Abdul-Masih et al.}
   \maketitle
%
%-------------------------------------------------------------------

\section{Introduction}
Through their ionizing fluxes, strong stellar winds and powerful explosions, massive stars play a crucial role in the star formation process and drive the chemical evolution of their host galaxies \citep[for a review see: ][]{Bresolin2008}.  Understanding how these massive stars form, evolve, and how they end their lives is thus of vital importance to our understanding of the universe in general. A common feature seen in massive star systems is a tendency to be found in close binary systems \citep[e.g., ][]{Sana2011}, and this can have a large effect on the evolution of the component stars.  In fact, it has been demonstrated that about 40 percent of all O-type stars will interact with a companion during their lifetimes \citep{Sana2012}, and about half of these interactions ($\sim$ 25 \% of all O-type stars) are expected result in an overcontact configuration \citep[e.g., ][]{Pols1994, Wellstein2001, deMink2007}. 

The overcontact phase represents a crossroad in the evolution of massive close binary systems. Depending on the nature and efficiency of the internal physical processes as well as the rate of mass transfer when the system initially comes into contact, this phase can lead to various exotic astronomical objects.  Often the overcontact phase is short lived, evolving on the thermal or dynamical timescale, meaning that it is rather rare to observe a system during this phase.  However if the conditions are right, stable overcontact systems that evolve on the nuclear timescale can form \citep{deMink2016, Mandel2016, Marchant2016, Menon2020}.  Here we focus on these stable systems.  While many factors affect the final fate of stable massive overcontact systems, one of the most important, and arguably least well constrained is internal mixing. If the mixing is efficient enough, the stars may enter the chemically homogeneous evolution regime \citep[CHE; ][]{Maeder1987}, which has several interesting evolutionary implications including the formation of long gamma-ray bursts \citep{Woosley2006,Yoon2006} and the production of ionizing photons in low metallicity star forming regions \citep{Szecsi2015} to name a few. Instead of expanding as they evolve, stars evolving via the CHE pathway may shrink.  This pathway has been proposed as a way to form gravitational wave progenitors \citep{deMink2016, Mandel2016, Marchant2016, duBuisson2020, Riley2020}.  If mixing is less efficient, then these systems will merge, potentially forming objects such as magnetic massive stars \citep{Schneider2019}, Be stars \citep{Shao2014}, LBVs \citep{Justham2014, Smith2018}, blue stragglers \citep{Eggen1989,Mateo1990} and more.  Thus, constraining these internal mixing processes is needed to accurately model the future evolution of massive overcontact systems and consequently their final fates. Mixing affects the internal structure and thus affects the stellar properties such as the temperature, radius, and luminosity.  If mixing extends to the surface, then it can also affect the surface chemical abundances.  By studying the temperature and surface abundances and comparing them with evolutionary models, we can constrain the degree of internal mixing during this phase \citep[e.g., ][]{deMink2009}.

In total, ten massive O+O overcontact binaries are currently known: V382 Cyg, TU Mus, MY Cam, UW CMa, HD 64315, LSS 3074, VFTS 066, VFTS 352, BAT99 126, and OGLE SMC-SC10 108086 \citep{Leung1978, Popper1978, Hilditch2005, Penny2008, Lorenzo2014, Lorenzo2017, Almeida2015, Howarth2015, Mahy2020a, Janssens2021}. Of these ten systems, six are located in the Milky Way, three in the Large Magellanic Cloud (LMC henceforth) and one in the Small Magellanic Cloud (SMC henceforth).  Despite its small size, this sample is fairly well distributed across the relevant parameter spaces of mass ratio, total system mass, fillout factor, period, component radii, metallicity and multiplicity, meaning that a full analysis of the complete sample will provide some much needed insights into the mixing mechanisms during this crucial phase. Note that the fillout factor $f$ is a measure of the degree to which a system is overfilling its roche lobes and has several different definitions in the literature.  In this study we use the definition first proposed by \citet{Mochnacki1972}:

   	\begin{equation}\label{eq:fillout_factor}
      f = \frac{\Omega_{n,1} - \Omega_n}{\Omega_{n,1} - \Omega_{n,2}} + 1,
   	\end{equation}

where $\Omega_{n,1}$ and $\Omega_{n,2}$ denote the normalized potential of the surface passing through L1 and L2 respectively, and $\Omega_n$ indicates the actual surface potential of the system.  In this definition, an overcontact system has a fillout factor $1 < f < 2$, with higher fillout factors corresponding to systems in deeper contact.  A fillout factor of exactly 1 implies a contact system where both components are exactly filling their roche lobes, while a fillout factor 2 implies that the system is at the limit of overflowing through the L2 Lagrangian point.

This series of papers aims to spectroscopically analyze O+O overcontact binaries to provide robust observational constraints on the internal mixing processes, and to better understand the properties and evolutionary outcomes of the shortest period massive binaries.  In this pilot study, we focus on three such systems in order to present our analysis techniques and investigate any initial trends or commonalities in the sample. 

The goal of this study is two-fold.  First, we aim to place constraints on the internal mixing processes during the overcontact phase of massive binary evolution, and second we investigate how the results of atmosphere fitting change when considering spherical geometry versus a more realistic three-dimensional geometry. In Sect. \ref{sec:sample} we discuss our sample and data reduction techniques.  Section \ref{sec:methods} details the two spectroscopic fitting techniques used, which both rely on the same underlying atmosphere models but differ in the adopted geometry: (i) assuming spherical geometry and (ii) using a realistic 3D mesh representation of the surface. In Sect. \ref{sec:results} we discuss the results of the two methods and in Sect. \ref{sec:discussion} we place our findings into an evolutionary context.  In this section, we also discuss how the results of the two analysis techniques compare to each other and we discuss systematic differences that arise.  Finally, Sect. \ref{sec:summary} summarizes our findings and discusses future prospects.

\section{Sample and Observations} \label{sec:sample}
\begin{figure*}
   \centering
   \includegraphics[width=1\linewidth]{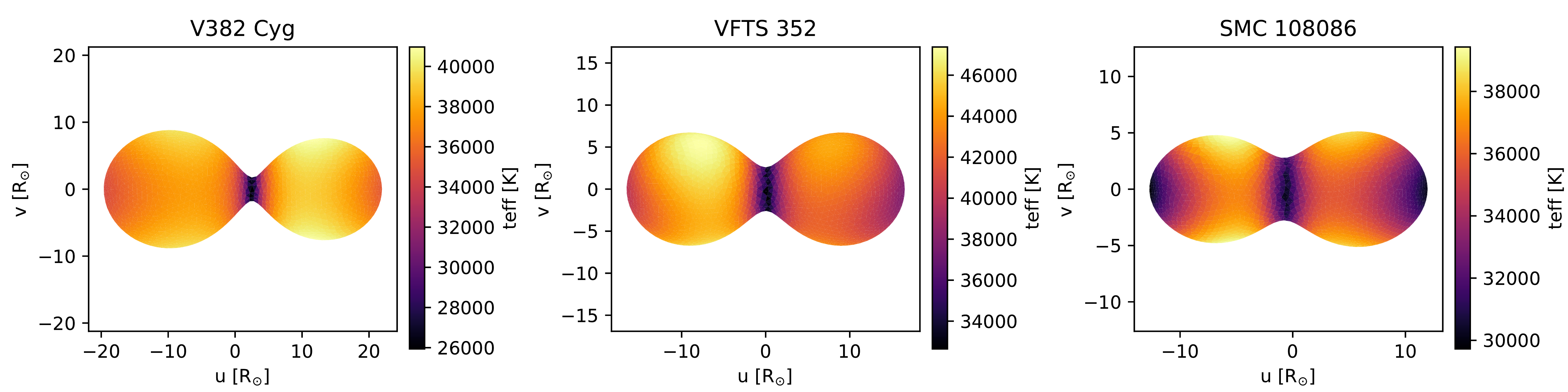}
      \caption{
      PHOEBE II mesh model of V382 Cyg (left), VFTS 352 (center) and SMC 108086 (right).  The mesh color represents the local effective temperature, with lighter colors representing higher temperatures and darker colors representing lower temperatures.  The systems have been inclined based on the the system inclinations as given in Table \ref{table:orb_sol}.
              }
         \label{teffs_mesh}
   \end{figure*}

Our sample consists of three massive overcontact systems at different metallicities.  These three objects were selected not only due to data availability and wavelength coverage, but also because they each probe different areas of the parameter space. The surface geometry and temperature structure for each system is shown in Fig. \ref{teffs_mesh}, and each system is discussed below.

\subsection{V382 Cyg}
V382 Cyg is located in the Milky Way and was the first massive overcontact binary identified \citep{Cester1978, Popper1978}. With a period of $\sim$1.886 days, the unequal masses of the component stars (26 + 19 \msol) as well as an observed period increase ($\sim$3.28 s per century) suggest that the system is undergoing conservative mass transfer from the less massive component to the more massive component at a rate of approximately $5.0 \times 10^{-6}$ \msun/yr  \citep{Degirmenci1999, Martins2017}.  This system was also studied spectroscopically by \citet{Martins2017} who found that the system showed baseline CNO surface abundances.  Modelling of the light curve indicates that the system has a fairly low fillout factor of 1.1, meaning that the components are only just over filling their Roche lobes \citep{Degirmenci1999, Martins2017}. 

Our data set for V382 Cyg consists of 89 well-phase-covered spectra collected over a two month period using the High-Efficiency and high-Resolution Mercator Echelle Spectrograph \citep[HERMES; ][]{Raskin2011} mounted on the Mercator Telescope in La Palma (HERMES Program 79; PI: M. Abdul-Masih).  The usable wavelength range of our observations spans from $\sim$4000 -- 9000 \AA\ with a spectral resolution of 85,000.  The data were reduced using the HERMES automatic pipeline resulting in data with similar signal-to-noise ratios per pixel at around 50 at each epoch.

\subsection{VFTS 352}
VFTS 352 is located in the Tarantula region of the LMC. Its sub-solar metallicity, short period ($\sim$1.124 days) and high component masses ($\sim$29 + 29 \msol) make this system a good candidate for the CHE pathway.  It was first discovered and characterized by \citet{Almeida2015} and was later studied spectroscopically by \citet{Abdul-Masih2019} and \citet{Mahy2020b}.  These studies showed that the combination of temperature and surface abundances could not be reproduced by single or binary evolutionary models, emphasizing our lack of understanding of the complex internal processes during the contact phase. With a fillout factor of 1.29 and a mass ratio of almost unity, the geometry of VFTS 352 is representative of a prototypical massive overcontact system, where a rapid phase of mass transfer has equalized the masses and the system is now thought to be in a long lasting slow case A phase evolving on the nuclear timescale.

For this current study, we use the spectroscopic data set from \citet{Abdul-Masih2019}, which consists of 8 far-UV spectra obtained with the Cosmic Origins Spectrograph (COS) on the Hubble Space Telescope (HST) under the auspices of program GO 13806 (PI: Sana) and 32 optical spectra obtained with the FLAMES-GIRAFFE spectrograph on the ESO VLT as part of the TMBM \citep[Tarantula Massive Binary Monitoring; PI: H. Sana; ESO programs: 090.D-0323 and 092.D-0136][]{Almeida2017}. For the purposes of this study, we utilize only the optical portion of the spectrum as the UV lines are much more affected by the wind parameters than the optical lines. The optical data spans from $\sim$3950 to 4550\AA, and is well phase-covered.  These data were reduced using the ESO CPL GIRAFFE pipeline v.2.12.1, resulting in data with similar signal-to-noise ratios per pixel at around 150 at each epoch.  Further details regarding the observational setup and data reduction can be found in Section 2 of \citet{Abdul-Masih2019}.
%The UV data set provides almost continuous wavelength coverage from $\sim$1130 to 1790\AA\ and is equally spaced in phase.  The data were reduced using the CALCOS 3.0 (2014 October 30) pipeline, which uses standard reduction techniques.  

\subsection{SMC 108086}
OGLE SMC-SC10 108086 (SMC 108086 henceforth) is located in the SMC and is the lowest metallicity massive overcontact system currently known.  This unequal mass system ($\sim$17 + 14 \msol) was photometrically characterized as a contact system by \citet{Hilditch2005} and the ephemeris and period were later updated by \citet{Pawlak2016}.  This system has both the shortest period ($\sim$0.883 days) and highest fillout factor ($\sim$1.7) of all currently known massive overcontact systems, making it an interesting candidate to study the internal mixing processes during this phase.

Our data set for SMC 108086 consists of 12 evenly phased spectra collected over a 6 month period (ESO program: 0103.D-0237; PI: M. Abdul-Masih) using the X-SHOOTER multi-wavelength spectrograph on the ESO VLT \citep{Vernet2011}.  Combining the three arms, the wavelength coverage spans from 3000 to 25\,000 \AA\ with spectral resolutions of 6700, 8900, and 5600 for the UVB, VIS and NIR arms respectively.  Each of the 12 epochs consists of one nodding cycle with 1415 s exposures in UVB, 1478 s in VIS and 1200 s in NIR.  These data were reduced following the standard procedure using the ESOReflex automated reduction pipeline \citep{Freudling2013}.  The resulting data at each individual epoch have similar signal-to-noise ratios per pixel at around 160 at $\sim$ 4800 \AA.

\subsection{Normalization}

Of the three systems in our sample, one (VFTS 352) has been spectroscopically analyzed in a previous study, so the data are already normalized \citep{Abdul-Masih2019}.  For the other two systems, we normalize by fitting a second order spline through a series of selected knot points that trace the continuum.  The wavelength of the knots are chosen by eye and a corresponding flux is calculated for each by taking the median of all flux points within a 1\AA\ region around the selected knot.  A spline is fit through these knots and the observed spectrum is divided by the resulting spline to obtain a final normalized spectrum.

\section{Spectral Analysis} \label{sec:methods}
As stated above, one of the goals of this study is to investigate the systematic differences between a spherically symmetric atmosphere fitting approach and a more realistic treatment of the three-dimensional surface geometry in the atmosphere fitting of overcontact systems.  We test this by fitting the observed spectra of each object in our sample using two distinct methods.  The first involves spectral disentangling followed by fitting each component individually using \fw \citep[a non-local thermodynamic equilibrium, NLTE henceforth, radiative transfer code designed to model the atmosphere and wind structure of massive stars;][]{Puls2005}.  The second method involves fitting each phase of the observed spectra using \spamms \citep[a spectroscopic patch model that models line profiles across the entire visible surface of a massive binary system;][]{Abdul-Masih2020a}.  In both cases, we assume smooth unclumped winds.

\subsection{Spherical Atmosphere Fitting}
Most of the codes currently available to spectroscopically model massive stars are one dimensional.  Massive stars require NLTE radiative transfer codes to accurately model their spectral lines, and the NLTE calculations are computationally expensive, making higher-dimensional modeling challenging.  For this reason, the components of massive binary systems are typically modeled separately as individual single spherically-symmetric stars.  This approach requires that the spectral contributions from each component first be separated from one another before atmosphere fitting can be performed.

\subsubsection{Spectral Disentangling}
	Before the spectra of each individual component of the binary can be fit with \textsc{fastwind}, the spectral contributions need to be disentangled \citep[for a review see][]{Pavlovski2010}. For contact binaries, the flux ratios are not constant over the orbit, so we require a code that can account the flux ratio at each epoch.  While there are several viable methods \citep[see e.g., ][]{Simon1994, Hadrava1995, Hadrava2009, Skoda2012}, we choose to use FDBinary \citep{Ilijic2004} to remain consistent with \citet{Abdul-Masih2019}.  FDBinary is a spectral disentangling code that works in Fourier space.  Based on the orbital solution and flux ratios at each epoch, FDBinary returns an individual disentangled spectrum for each component.  To obtain the light ratios at each phase, we model the system using the Wilson-Devinney-like code \textsc{phoebe ii} \citep[\phoebe hereafter; ][]{Prsa2016, Horvat2018, Jones2020, Conroy2020} and calculate the flux of the primary and secondary components separately.  The orbital solutions used for the \phoebe modeling and spectral disentangling for each system are given in Table \ref{table:orb_sol}.

\begin{table}
\caption{Orbital solutions for V382 Cyg \citep{Martins2017}, VFTS 352 \citep{Almeida2015}, and SMC 108086 \citep{Hilditch2005}.  In the case of SMC 108086, we use the updated $P_\mathrm{orb}$ and $T_\mathrm{0}$ from \citet{Pawlak2016}.  Note that $T_\mathrm{0}$ refers to the time of superior conjunction and that all systems are assumed to be circular.}
\centering 
\begin{tabular}{cccc}
\hline\hline
 & V382 Cyg & VFTS 352 & SMC 108086 \\
\hline
$P_\mathrm{orb}$ \ (day) & $1.885545$ & $1.1241452$ & $0.8830987$\\ 
$T_\mathrm{0}$ \ (HJD)   & $2456527.020$ & $2455261.119$ & $2457000.7307$\\
$M_1$ (\msun)            & 26.1 $\pm$ 0.4 & 28.63 $\pm$ 0.3 & 16.9 $\pm$ 1.2 \\
$M_2$ (\msun)            & 19.0 $\pm$ 0.3 & 28.85 $\pm$ 0.3 & 14.3 $\pm$ 1.7 \\
$q$ ($M_1/M_2$)          & 1.376 $\pm$ 0.009 & 0.99 $\pm$ 0.01 & 1.183 $\pm$ 0.080 \\
$R_1$ (\rsun)            & 9.4 $\pm$ 0.2 & 7.22 $\pm$ 0.02 & 5.7 $\pm$ 0.2 \\
$R_2$ (\rsun)            & 8.7 $\pm$ 0.2 & 7.25 $\pm$ 0.02 & 5.3 $\pm$ 0.2 \\
$f$                      & 1.1 & 1.29 & 1.7 \\
$K_\mathrm{1}$ (\kms )   & 257 & $327.8$ & 317 \\
$K_\mathrm{2}$ (\kms )   & 354 & $324.5$ & 375 \\
$i$ ($^\circ$)           & $85$ & $55.6$ & $82.8$ \\

%$e$                      & 0 (fixed)  \\
%$\omega$                 & -90 (fixed) \\
 \hline
\end{tabular}
\label{table:orb_sol}
\end{table}

\subsubsection{FASTWIND Fitting}
	Now that the spectral contributions of each component are separated, we can fit each one with \textsc{fastwind} \citep[v. 10.3][]{Puls2005, Sundqvist2018a}.  \fw itself does not have fitting capabilities, so we use a Genetic Algorithm optimization routine wrapped around \fw to perform the fitting. This method has been used and discussed at length in several previous works \citep[e.g. ][]{Mokiem2005, Mokiem2006, Mokiem2007, Tramper2011, Tramper2014, Ramirez-Agudelo2017, Abdul-Masih2019}.  In this work, we present pyGA\footnote{https://github.com/MichaelAbdul-Masih/pyGA}, a new Genetic Algorithm written in Python based on the exploration methods introduced in the FORTRAN code \textsc{pikaia} \citep{Charbonneau1995}. 
	
	As with other Genetic Algorithms, pyGA functions under the principles of "survival of the fittest" \citep{Darwin1859}.  An initial population of models is created with each individual having a randomly assigned combination of parameters within the user defined parameter space.  These parameters are analogous to genes, and the combination of parameters can be thought of as a chromosome.  Individuals with the most favorable set of genes have higher chances of passing on their genetic material to the next generation.  In this case, models that are better able to reproduce the observed spectra (i.e. models with a higher fitness metric) are given a higher weight when choosing the parents for the next generation. The fitness metric is closely related to the chi square and is calculated as follows:
	\begin{equation}
      \mathrm{Fitness} \equiv \left( \sum_{i}^{N} \chi_{\mathrm{red}, i}^2 \right) ^ {-1},
   	\end{equation}
    where $N$ represents the total number of spectral lines being fitted over and $\chi_{\mathrm{red}, i}^2$ represents the reduced chi square of the $i^\mathrm{th}$ spectral line \citep{Mokiem2005}.  The next generation is created by combining the chromosomes from the parents using genetic concepts such as crossovers and mutations.  After several generations, the algorithm will converge on the best fit solution.   pyGA has been written such that it can be easily applied to a variety of problems; a beta version of the code has already been applied to light curve fitting as well \citep{Sekaran2020}.
	
	In this study, we setup the Genetic Algorithm (GA henceforth) mirroring the setup in \citet{Abdul-Masih2019}. As in \citet{Abdul-Masih2019}, we perform an 11 parameter optimization, fitting the stellar (effective temperature, surface gravity and rotation rate) and wind parameters (mass loss rate, beta parameter and terminal wind speed) as well as the surface abundances (helium, carbon, nitrogen, oxygen and silicon) for each component of the contact systems in our sample. For the purposes of this study, the abundance of helium is given by:
	\begin{equation}
      Y_\mathrm{He} = \frac{N_\mathrm{He}}{N_\mathrm{H}},
   	\end{equation}
	 where $N_\textrm{He}$ and $N_\textrm{H}$ are the number densities of helium and hydrogen respectively.  Note that here we refer to $Y_\mathrm{He}$ as the number density not mass density. The abundances of the other elements are given by:
	\begin{equation}
      \varepsilon_\mathrm{X} = \log \frac{N_\mathrm{X}}{N_\mathrm{H}} + 12,
   	\end{equation}
     where $N_\textrm{X}$ and $N_\textrm{H}$ are the number densities of the given element and hydrogen respectively. 
     
     For each component, we fix the radius to those given in Table \ref{table:orb_sol}.  We simultaneously fit about 25 spectral lines (and blends) including species of hydrogen, helium, carbon, nitrogen, oxygen and silicon.  A list of diagnostic lines and wavelength ranges that are used for V382 Cyg and SMC 108086 can be found in Table \ref{table:line_list}.
     
     To remain consistent with \citet{Abdul-Masih2019}, the error calculation is conducted in the same way.  This is done by first normalizing the chi square values such that the model with the lowest chi square satisfies $\chi_{\mathrm{red}}^2 = 1$, which assumes that this model provides a satisfactory fit to the data.  We then calculate the probability ($P$) that the deviations in the normalized $\chi_{\mathrm{red}}^2$ of each model is not caused by statistical fluctuations.  The probability is given by:
     \begin{equation}
      P=1-\Gamma\left(\frac{\chi^2}{2}, \frac{\nu}{2}\right),
   	\end{equation}
     where $\Gamma$ is the incomplete gamma function and $\nu$ is the degrees of freedom.  Models that satisfy $P\ge 0.05$ (representing a 95\% confidence interval) are considered part of the family of acceptable solutions. Thus the errors on the stellar and wind parameters are given as the ranges spanned by all models which are part of the family of acceptable solutions.  These regions is indicated in Figures \ref{ga_fitness_plots-v382_a}, \ref{ga_fitness_plots-v382_b}, \ref{ga_fitness_plots-smc_a} and \ref{ga_fitness_plots-smc_b} in Appendix \ref{appendix1} by the shaded blue regions in each parameter plot.
     
     While our fitting method does not include luminosity directly as a fitting parameter, we can calculate it outside of the GA given the best-fit parameters.  Typically, the luminosities for massive overcontact systems are calculated via the Stefan-Boltzmann Law, so to remain consistent with the literature, we do the same.  This relation implicitly assumes that the stars are spherical, which is already assumed in the GA method, so this does not add any additional assumptions.

	\begin{table*}
\caption{Summary of the diagnostic line list. The main identifier is given in Col.~1 although it should be noted that some of these lines contain blends. The adopted fitting ranges per line are provided in Cols.~2 and 3 for V382 Cyg and SMC 108086 respectively.  All of the individual line components that are included in the calculations of each line are indicated in Col.~4.}
\centering 
\begin{tabular}{ccccc}
\hline\hline
Line Identifier &  \multicolumn{2}{c}{$\lambda$ Fitting Range (\AA)} & Line Components\\
 & V382 Cyg & SMC 108086 & \\
\hline
H$\delta$ 			& 4082.0 -- 4111.8  &  4085.1 -- 4117.1	& H \textsc{i} 4101.734, He \textsc{ii} 4101.198, N \textsc{iii} 4097.35, 4103.43, Si \textsc{iv} 4088.862\\
H$\gamma$ 			& 4327.6 -- 4358.8	&  4329.2 -- 4359.4 & H \textsc{i} 4340.472, He \textsc{ii} 4339.891, Si \textsc{iv} 4328.177 \\
H$\beta$ 			& 4846.8 -- 4874.0	&  4849.8 -- 4879.7 & H \textsc{i} 4861.35, He \textsc{ii} 4861.35 \\
H$\alpha$ 			& 6546.6 -- 6578.5	&  6534.4 -- 6594.7 & H \textsc{i} 6562.79, He \textsc{ii} 6562.79 \\
He \textsc{i} 4121 	& 4112.4 -- 4126.4  &  4117.0 -- 4129.2	& He \textsc{i} 4120.815, Si \textsc{iv} 4116.104\\
He \textsc{i} 4143 	& 4137.2 -- 4152.9  &  4137.4 -- 4156.4	& He \textsc{i} 4143.761\\
He \textsc{i} 4387 	& 4382.5 -- 4395.6  &  4381.9 -- 4399.1	& He \textsc{i} 4387.930\\
He \textsc{i} 4471 	& 4462.0 -- 4479.8  &  4464.4 -- 4485.7	& He \textsc{i} 4471.480\\
He \textsc{i} 4713 	& 4704.7 -- 4720.5  &  4706.2 -- 4726.0	& He \textsc{i} 4713.145\\
He \textsc{i} 4922 	& 4912.7 -- 4931.7  &  4914.7 -- 4936.3	& He \textsc{i} 4921.931\\
He \textsc{i} 5016 	& 5007.8 -- 5023.8  &  5008.6 -- 5027.5	& He \textsc{i} 5015.678\\
He \textsc{i} 5875 	& 5863.2 -- 5883.9  &  ---	            & He \textsc{i} 5875.621\\
He \textsc{i} 6678 	& 6667.0 -- 6696.1  &  6664.7 -- 6703.8	& He \textsc{i} 6678.151, He \textsc{ii} 6685.046\\
He \textsc{ii} 4200 & 4189.8 -- 4209.8  &  4188.5 -- 4214.8	& He \textsc{ii} 4199.870, N \textsc{iii} 4195.76, 4200.10\\
He \textsc{ii} 4541 & 4530.2 -- 4553.2  &  4529.9 -- 4560.6	& He \textsc{ii} 4541.625\\
He \textsc{ii} 4686 & 4675.8 -- 4696.1  &  4676.0 -- 4702.0	& He \textsc{ii} 4685.742\\
He \textsc{ii} 5411 & 5397.5 -- 5428.0  &  5399.9 -- 5432.1	& He \textsc{ii} 5411.554\\
C \textsc{iii} 4069 & 4064.5 -- 4078.7  &  4065.7 -- 4081.8 & C \textsc{iii} 4067.940, 4068.916, 4068.916, 4070.260\\
C \textsc{iii} 4187 & 4177.5 -- 4189.6  &  4180.6 -- 4192.8 & C \textsc{iii} 4186.900\\
C \textsc{iii} 4650 & 4647.1 -- 4657.1  &  4645.4 -- 4659.8	& C \textsc{iii} 4647.418, 4650.246, 4651.473\\
C \textsc{iv} 5801  & 5788.8 -- 5823.7	&  5793.2 -- 5828.1 & C \textsc{iv} 5801.33, 5811.98\\
N \textsc{iii} 4379 & 4375.6 -- 4382.3	&  4378.7 -- 4383.4 & N \textsc{iii} 4378.93, 4379.11\\
N \textsc{iii} 4515 & 4508.7 -- 4521.2	&  4512.1 -- 4524.6 & N \textsc{iii} 4514.854, 4510.965, 4510.885, 4518.143\\
N \textsc{iii} 4640 & 4629.4 -- 4645.5	&  4632.8 -- 4645.8 & N \textsc{iii} 4634.122, 4640.641, 4641.850 \\
N \textsc{iv} 4058  & 4052.7 -- 4063.8  &  4055.8 -- 4066.9	& N \textsc{iv} 4057.76\\
O \textsc{iii} 5592  & 5581.5 -- 5602.8	&  5585.7 -- 5607.0 & O \textsc{iii} 5592.252\\
 \hline
\end{tabular}
\label{table:line_list}
%\label{table:line_list}
\end{table*}

\subsection{3D Surface Geometry Atmosphere Fitting}	
Assuming it is in hydrostatic equilibrium in the co-rotating frame of the binary, the surface geometry of an overcontact system can be approximated using the Roche formalism, which states that the system is bound by an equipotential surface.  This implies that across the surface there can be a range of effective surface gravities and from the von Zeipel theorem, a range of effective temperatures \citep{vonZeipel1924}.  Thus, depending on the degree of surface distortion and the inclination of the system, the observed spectral lines will differ from a model assuming spherical geometry \citep{Abdul-Masih2020a}.  To account for these three-dimensional effects, we use the \spamms code to model these overcontact systems.

\spamms is a spectral analysis tool designed for distorted massive stars.  This code combines the Wilson-Devinney-like binary modeling code \phoebe with the NLTE radiative transfer code \fw to compute a patch model for massive systems in various configurations. Given a binary solution and effective temperatures of the components, \spamms first uses \phoebe to compute a mesh that represents the surface of the stars in the system and then it populates these mesh points with local parameters.  Since the local temperature profile across the surface of a distorted star is not constant, the effective temperature represents an intensity weighted average temperature across the surface.  Using the von Zeipel theorem and the provided effective temperature, \phoebe calculates the local temperature at each point \citep[for a full description of this process, see section 5.1 of ][]{Prsa2016}.  For these computations, we use the standard values recommended by \phoebe for massive hot stars, namely: a bolometric gravity brightening coefficient of 1.0, a reflection coefficient of 1.0, a logarithmic limb darkening prescription and blackbody atmospheres.  Based on the local temperature, surface gravity, and radius, \spamms then assigns \fw emergent intensity line profiles to each mesh point.  Finally, \spamms integrates over the visible surface to return a line profile for the entire system at the given phase and orientation.  While the patch model still uses 1D model atmospheres for each patch, it better accounts for the surface geometry as well as the surface gravity and temperature structure across the surface. This method can not only handle complex non-spherical geometries, but it also accounts for the relative light contributions of the stars in the system and allows both stars to be modeled simultaneously.

Because the surface geometry is taken into account and the input grid assumes a \citet{Vink2001} mass-loss prescription \citep[see ][ for details on the wind implementation in \spamms and Sect. \ref{sec:discussion_3d_1d} of this paper for a further discussion on the consequences of the associated assumptions]{Abdul-Masih2020a}, fitting with \spamms requires fewer free parameters than fitting with \textsc{fastwind}.  For this reason, we choose to fit using a grid-search chi-square minimization routine.  Using this method, we optimize for six parameters: the effective temperature of the primary and secondary, and the surface abundances of helium, carbon, nitrogen and oxygen for the system as a whole.  We do not separate the surface abundance measurements of the two components because, as shown in \citet{Abdul-Masih2019} and Sect. \ref{sec:results_GA} of this paper, the surface abundances appear consistent between the components in overcontact systems.  The geometry of the system is constrained from the photometric orbital solutions, given in Table \ref{table:orb_sol}, and all observational epochs are fit simultaneously resulting in a global best-fit solution for both the primary and secondary.  Note that our fitting process does not include the surface gravity as this is already accounted for in the geometry of the mesh model, which requires the masses and equivalent spherical radii as input parameters.

Errors for the fits are determined via chi square statistics.  For each fitting parameter, we calculate the minimum chi square value per grid point.  Using these minimum values, we fit a cubic spline and determine the minimum of the spline fit.  Using this global chi square minimum, we then calculate the chi square value corresponding to a 1 sigma error ($\chi_{1\sigma}$) using:

\begin{equation}
    \chi_{1\sigma} = \chi_\mathrm{min} \left(1 + \sqrt{\frac{2}{n_\mathrm{dof}}}\right),
\end{equation}

where $\chi_\mathrm{min}$ is the global chi square minimum as determined by the cubic spline fit and $n_\mathrm{dof}$ is the number of degrees of freedom \citep{Tkachenko2015}.  The resulting $\chi_{1\sigma}$ is compared to the cubic chi square spline fit and confidence intervals are determined based on where the chi square spline fit is below the $\chi_{1\sigma}$ threshold.

Preliminary fits indicated that synchronous rotation produced lines that were too narrow for two of the systems in the sample.  To account for possible additional broadening mechanisms such as macroturbulence or asynchronous rotation, we introduce an asynchronicity parameter $F$ to the patch model for each star in the system.  We define this parameter as $F = v_\mathrm{rot}/v_\mathrm{tl}$ where $v_\mathrm{rot}$ is the rotational velocity and $v_\mathrm{tl}$ is the rotation rate assuming tidal locking.  Since \phoebe does not allow asynchronous rotation in overcontact systems, this is implemented outside of \phoebe as a perturbation to the radial velocities of each mesh point.  After the \phoebe model is computed, we subtract the radial velocity of the component from the radial velocity of each mesh grid point associated to that component.  This produces radial velocities which are centered about the axis of rotation (as opposed to the orbit).  We then multiply all radial velocities across the component's mesh by the asychronicity parameter and add the component's radial velocity back in.  In reality, asynchronous rotation would alter the mesh geometry, but it is important to note that our current implementation does not change the geometry of the mesh.  This process mimics a higher rotation rate, however in this case we are using the asynchronicity parameter as a proxy to estimate the additional broadening needed to reproduce the observed spectra.  In addition, because \spamms calculates the line profiles for the entire visible surface simultaneously, by implementing this asynchronicity parameter, we can investigate the extra broadening in each component separately.  This will allow us to determine whether additional broadening is needed in both components or only one, and whether the additional broadening is correlated with any of the other parameters.

The fitting itself is done in two stages.  First we simultaneously constrain the temperature and  asynchronicity parameter for the primary and secondary using the most sensitive temperature diagnostics in our sample, namely \hea~\l4471 and \heb~\l4541.  Using the best fit solution from this initial stage, we then constrain the CNO and helium surface abundances by fitting lines from our line list given in Table \ref{table:line_list} that contain only CNO elements and helium.  For the surface abundance fitting stage, we calculate a finer \spamms input abundance grid containing seven helium steps (0.06, 0.08, 0.10, 0.125, 0.15, 0.175 and 0.20) and 13 CNO abundance steps ranging from 6.0 to 9.0 inclusive in steps of 0.25.  This is done to ensure that the abundance errors are not limited by our coarse grid size.  Finally, we compute a model with the best fit parameters to ensure that the rest of the lines in the sample are well reproduced.

As with the GA, \spamms does not directly include luminosity in the fitting, however since \spamms accounts for the 3D geometry, applying the Stefan-Boltzmann law as is may not be valid.  Instead, we use the local parameters across the mesh to compute the luminosity contribution of each patch and then sum them together. The luminosity contribution of each patch is calculated with a modified Stefan-Boltzmann relation, which replaces the spherical surface area term ($4\pi R^2$) with the surface area, $a$ of the given patch.  Thus, the luminosity is given by:
\begin{equation}
    L_\mathrm{3D} = \sum_{i}^{N}(a_i \sigma_\mathrm{SB} T_{\mathrm{eff}, i}^4),
\end{equation}
 where $i$ is the specific patch in question, $N$ is the total number of patches across the surface, $a_i$ and $T_{\mathrm{eff}, i}$ are the area and local temperature of the patch, respectively and $\sigma_\mathrm{SB}$ is the Stefan-Boltzmann constant.

\section{Results}  \label{sec:results}

\begin{table*}
\caption{Results of the spherical GA analysis for the primary and secondary components of the systems in our sample.  Column 2 indicates the units, when applicable of the parameter in column 1.  When a parameter is expressed in log scale, the units are indicated with square brackets instead of parentheses.  The analysis of VFTS 352 was not performed in this study, but is given here for comparison \citep{Abdul-Masih2019}. The parameters corresponding to the model with the lowest $\chi^2$ are given and the 95$\%$ confidence interval for each is indicated.  Additionally, while the luminosity is not an optimized parameter in the GA analysis, we include it here for convenience. Note that the mass loss rates refer to unclumped winds. }
\centering 

\setlength{\extrarowheight}{6pt}
\begin{tabular}{ccccccccccc}
\hline\hline
 & & & \multicolumn{2}{c}{V382 Cyg} & & \multicolumn{2}{c}{VFTS 352} & & \multicolumn{2}{c}{SMC 108086} \\
 & & & primary & secondary & & primary & secondary & & primary & secondary \\
\hline
$T_\mathrm{eff}$ &  $\left(\mathrm{K}\right)$ &  & $36254^{+314}_{-80}$ & $37048^{+41}_{-3}$ &  & $44200^{+1350}_{-1350}$ & $40750^{+800}_{-150}$ &  & $33626^{+916}_{-1498}$ & $34249^{+1500}_{-1101}$\\
%log $g$ &  & $3.72^{+0.1}_{-0.1}$ & $3.88^{+0.03}_{-0.13}$ &  & $4.14^{+0.1}_{-0.05}$ & $3.9^{+0.1}_{-0.1}$ &  & $4.08^{+0.14}_{-0.19}$ & $3.75^{+0.32}_{-0.13}$\\
log $g$ & $\left[\mathrm{cm\ s}^{-2}\right]$ &  & $3.85^{+0.1}_{-0.1}$ & $3.96^{+0.03}_{-0.13}$ &  & $4.14^{+0.1}_{-0.05}$ & $3.9^{+0.1}_{-0.1}$ &  & $4.24^{+0.14}_{-0.19}$ & $4.00^{+0.32}_{-0.13}$\\
$\log \dot{M}$ & $\left[\mathrm{M}_\odot\,\mathrm{yr}^{-1}\right]$ &  & $-6.18^{+0.15}_{-0.09}$ & $-6.16^{+0.04}_{-0.15}$ &  & $-7.1^{+0.15}_{-0.15}$ & $-7.05^{+0.05}_{-0.45}$ &  & $-7.97^{+1.52}_{-0.53}$ & $-6.63^{+0.61}_{-1.86}$\\
$\beta$ & & & $1.79^{+0.14}_{-0.3}$ & $1.93^{+0.06}_{-0.25}$ &  & $3.3^{+0.7}_{-1.0}$ & $1.55^{+1.4}_{-0.1}$ &  & $0.64^{+3.35}_{-0.14}$ & $0.98^{+2.99}_{-0.48}$\\
$v_\infty$ & $\left(\mathrm{km\ s}^{-1}\right)$ &  & $3454^{+539}_{-1308}$ & $3637^{+358}_{-1190}$ &  & $2300^{+600}_{-300}$ & $2600^{+1400}_{-400}$ &  & $1686^{+2201}_{-684}$ & $1460^{+2531}_{-460}$\\
$v$ sin $i$ & $\left(\mathrm{km\ s}^{-1}\right)$ &  & $350.88^{+27.7}_{-6.01}$ & $315.86^{+6.59}_{-30.0}$ &  & $268.0^{+16.0}_{-28.0}$ & $296.0^{+14.0}_{-18.0}$ &  & $461.87^{+19.91}_{-41.77}$ & $406.72^{+57.81}_{-41.21}$\\
$Y_\mathrm{He}$ & &  & $0.09^{+0.01}_{-0.03}$ & $0.09^{+0.02}_{-0.0}$ &  & $0.1^{+0.03}_{-0.01}$ & $0.08^{+0.02}_{-0.01}$ &  & $0.09^{+0.04}_{-0.02}$ & $0.07^{+0.05}_{-0.02}$\\
$\varepsilon_\mathrm{C}$ & &  & $8.13^{+0.26}_{-0.05}$ & $8.18^{+0.1}_{-0.08}$ &  & $7.7^{+0.3}_{-0.3}$ & $7.25^{+0.3}_{-0.2}$ &  & $6.69^{+0.47}_{-0.67}$ & $6.85^{+0.37}_{-0.85}$\\
$\varepsilon_\mathrm{N}$ & &  & $7.55^{+0.23}_{-0.49}$ & $7.62^{+0.27}_{-0.73}$ &  & $6.4^{+1.15}_{-0.3}$ & $6.2^{+1.15}_{-0.2}$ &  & $7.42^{+0.46}_{-1.36}$ & $7.3^{+0.57}_{-1.29}$\\
$\varepsilon_\mathrm{O}$ & &  & $8.38^{+0.24}_{-0.54}$ & $8.33^{+0.59}_{-0.39}$ &  & $8.35^{+0.3}_{-0.9}$ & $8.0^{+1.0}_{-0.55}$ &  & $7.39^{+1.08}_{-1.38}$ & $6.7^{+2.26}_{-0.7}$\\
$\varepsilon_\mathrm{Si}$ & &  & $7.12^{+0.04}_{-0.64}$ & $6.91^{+0.27}_{-0.85}$ &  & $6.95^{+0.5}_{-0.95}$ & $6.5^{+0.65}_{-0.5}$ &  & $6.0^{+0.43}_{-0.0}$ & $6.0^{+0.65}_{-0.0}$\\
\hline
$\log(L_\mathrm{SB}/L_\odot)$ & &  & $5.14^{+0.02}_{-0.02}$ & $5.11^{+0.02}_{-0.02}$ &  & $5.25^{+0.05}_{-0.05}$ & $5.11^{+0.03}_{-0.01}$ &  & $4.57^{+0.06}_{-0.08}$ & $4.54^{+0.08}_{-0.06}$\\
$\log(L_\mathrm{dist}/L_\odot)$ & &  & $5.09^{+0.05}_{-0.05}$ & $5.10^{+0.05}_{-0.05}$ &  & $5.24^{+0.04}_{-0.04}$ & $5.16^{+0.04}_{-0.04}$ &  & $4.60^{+0.03}_{-0.03}$ & $4.66^{+0.03}_{-0.03}$\\

 \hline
\end{tabular}
\label{table:GA_results}
\end{table*}

\begin{table}
\caption{Reference baseline surface abundances for helium, carbon, nitrogen and oxygen for the Milky way, LMC and SMC.  This table is adapted from Tables 1 and 2 from \cite{Brott2011a}.}
\centering 

\setlength{\extrarowheight}{6pt}
\begin{tabular}{cccccc}
\hline\hline
 &  & & Milky Way & LMC & SMC \\
\hline
$Y_\mathrm{He}$ &  &  & 0.0907 & 0.0867 & 0.0842\\
$\varepsilon_\mathrm{C}$ &  &  & 8.13 & 7.75 & 7.37\\
$\varepsilon_\mathrm{N}$ &  &  & 7.64 & 6.90 & 6.50\\
$\varepsilon_\mathrm{O}$ &  &  & 8.55 & 8.35 & 7.98\\

 \hline
\end{tabular}
\label{table:baseline_abundances}
\end{table}

\begin{table}
\caption{Results of the \spamms analysis for each of the systems in our sample.  The parameters corresponding to the minimum interpolated $\chi^2$ are given and the confidence interval for each is indicated.  Cases where the confidence interval reaches the edge of the explored parameter space are indicated by parentheses.  While not included in the fit, the calculated luminosities are provided here for convenience.}
\centering 

\setlength{\extrarowheight}{6pt}
\begin{tabular}{cccccc}
\hline\hline
 &  & & V382 Cyg & VFTS 352 & SMC 108086 \\
\hline
$T_\mathrm{eff, 1}$ (K) &  &  & $37200^{+690}_{-720}$ & $44150^{+1100}_{-1200}$ & $36000^{+685}_{-660}$\\
$T_\mathrm{eff, 2}$ (K) &  &  & $38250^{+725}_{-750}$ & $41450^{+1170}_{-800}$ & $35200^{+635}_{-720}$\\
$F_1$ &  &  & $1.2^{+0.09}_{-0.1}$ & --- & ---\\
$F_2$ &  &  & --- & --- & $1.23^{+0.12}_{-0.1}$\\
$Y_\mathrm{He}$ &  &  & $0.06^{+0.01}_{(-0.0)}$ & $0.07^{+0.01}_{(-0.01)}$ & $0.09^{+0.03}_{(-0.03)}$\\
$\varepsilon_\mathrm{C}$ &  &  & $8.84^{+0.14}_{-0.15}$ & $7.47^{+0.31}_{-0.8}$ & $7.36^{+0.48}_{(-1.36)}$\\
$\varepsilon_\mathrm{N}$ &  &  & $7.17^{+0.41}_{-0.84}$ & $7.29^{+0.47}_{-1.12}$ & $7.6^{+0.54}_{(-1.6)}$\\
$\varepsilon_\mathrm{O}$ &  &  & $8.78^{(+0.22)}_{-0.4}$ & $7.5^{(+1.5)}_{(-1.5)}$ & $7.82^{+0.95}_{(-1.82)}$\\
\hline
$\log(L_\mathrm{3D, 1}/L_\odot)$ &  &  & $5.20^{+0.03}_{-0.04}$ & $5.24^{+0.05}_{-0.04}$ & $4.62^{+0.03}_{-0.04}$\\
$\log(L_\mathrm{3D, 2}/L_\odot)$ &  &  & $5.12^{+0.04}_{-0.03}$ & $5.14^{+0.05}_{-0.03}$ & $4.64^{+0.03}_{-0.04}$\\

 \hline
\end{tabular}
\label{table:spamms_results}
\end{table}

\subsection{Spherical Atmosphere Fitting Results} \label{sec:results_GA}
We perform the GA analysis for both the primary and secondary components of V382 Cyg and SMC 108086.  The results are given in Table \ref{table:GA_results}, and the $\chi^2$ plots per parameter and individual line profile fits can be found in Appendix \ref{appendix1}.  Since this same analysis has already been performed for VFTS 352 by \citet{Abdul-Masih2019}, we do not repeat the analysis, however for convenience and for comparison purposes, the results of that study are also given in Table \ref{table:GA_results}.    It is important to note that all models in the family of acceptable solutions are statistically equivalent, and for this reason, the parameter ranges are more important and informative than the model with the lowest chi square.

\subsubsection{V382 Cyg}
The effective temperature ranges of the primary and secondary are similar with values between $\sim$36\,175--36\,560 K and 37\,045--37\,090 K respectively, however the error bars for the secondary appear to be underestimated, with a more realistic range being closer to $\sim$36900--37200 K.  The derived surface gravities in units of \cmss\ of $\log g \approx$ 3.7--4.1 are consistent with predictions for main-sequence O-type dwarfs. These surface gravities have been corrected for centrifugal rotation effects following the procedure outlined in \citet{Repolust2004}. The measured projected rotational velocities of $\sim$350 and 315 \kms\ (for the primary and secondary respectively) are higher than expected when assuming tidal locking, which gives values of 251 and 232 \kms, however macroturbulent broadening effects are not accounted for in our fitting and may contribute to the observed additional broadening.  The derived mass loss rates in \msolyr are both in the range of $\log M_\odot \approx -$6.0 to $-$6.3, which is lower than predicted values using the \citet{Vink2001} prescription by about half a dex, and higher than predicted using the \citet{Bjorklund2021} prescription by almost a dex.  The beta parameter and terminal wind speed are both higher than expected when compared with typical values for O-type stars \citep[see e.g. ][]{Castor1975, Puls2008}, but without diagnostics in the UV, these are difficult to properly constrain.  The helium surface abundances are in the range of 0.06--0.11.  Both components show similar carbon, nitrogen and oxygen surface abundances of 8.1--8.4, 7.0--7.9, and 7.8--8.9 respectively.  Based on the mean radii from the photometric solutions and the effective temperatures measured in this work, we calculate luminosities of the primary and secondary to be $\log(L_\mathrm{SB}/L_\odot) \approx $ 5.12--5.16 and 5.09--5.13 respectively.  

Comparing our results with those of \citet{Martins2017}, we find a fairly good agreement overall.  The effective temperatures of the two components match very well as does the difference between the effective temperatures of the primary and secondary components.  Additionally, the surface gravities and luminosities also match almost exactly.  Interestingly, \citet{Martins2017} measures a much lower rotational velocity for both components that is consistent with tidal locking.  The surface abundances of carbon and oxygen are also in good agreement for the primary, but we measure a lower nitrogen abundance than \citet{Martins2017}.  Additionally, while \citet{Martins2017} finds a significant enhancement of carbon, nitrogen, and oxygen surface abundances in the secondary, our measurements do not indicate any such discrepancy between the two components.

\subsubsection{SMC 108086}
Like with V382 Cyg, the temperatures for both components of SMC 108086 are also very similar with ranges of $\sim$32\,125--34\,550 K and 33\,150--35\,750 K for the primary and secondary respectively.  The log of the surface gravities of the components are slightly different with ranges of 4.1--4.4 for the primary and 3.9--4.3 for the secondary, however both are within the predicted range.  Again, the measured projected rotational velocities (420--480 \kms\ and 370--470 \kms) are again higher than expected when assuming tidal locking (326 and 303 \kms).  The log of the measured mass loss rates both range from $-$8.5 to $-$6.0 which matches within errors of the predicted \citet{Vink2001} values of $-$8.06 and $-$8.03, but are higher than the predicted \citet{Bjorklund2021} values of $-$9.1 and $-$9.3.  As with V382 Cyg, the beta parameter and terminal wind speed are fairly unconstrained since the wavelength range does not extend down into the UV.  For both components, the surface abundances of helium range from 0.05--0.13, carbon from 6.0--7.2 and nitrogen from 6.0--7.8.  The oxygen surface abundances on the other hand are completely unconstrained, however this is not unexpected as our line list only contains one weak oxygen line.  The calculated luminosities of the both components are $\log(L_\mathrm{SB}/L_\odot) \approx $ 4.48--4.63.

In the case of both V382 Cyg and SMC 108086, previous studies have provided estimates for the temperatures and luminosities of the two components \citep[][ respectively]{Degirmenci1999, Hilditch2005} that do not agree with our measurements within errors.  It should be noted, however that both of these studies are photometric studies and neither derive these parameters through spectral fitting. Light curve fitting is much more sensitive to the ratio of the temperatures than their absolute values so an anchor temperature is often assumed for one of the components.  In both \citet{Degirmenci1999} and \citet{Hilditch2005}, the effective temperature of the primary was estimated based on the spectral type and was used as the temperature anchor.  In the case of \citet{Hilditch2005}, the luminosities of both components were calculated via the Stefan-Boltzmann law using the temperatures and radii determined from the photometric fit.  In \citet{Degirmenci1999} on the other hand, the luminosity was left as a free parameter in the photometric solution, however again the temperature of the primary was anchored based on spectral type.  In both cases, changes in the measured effective temperature will result in changes to the luminosity.  Since we perform a full spectroscopic analysis, our temperature measurements, and thus our resulting luminosity measurements, are more robust and reliable than those presented in these previous studies.  

\subsection{3D Surface Geometry Atmosphere Fitting Results}
We perform the 2 stage \spamms fitting for each object in our sample.  The results are provided in Table \ref{table:spamms_results} and the $\chi^2$ plots per parameter can be found in Appendix \ref{appendix2}.

\subsubsection{V382 Cyg}
The temperatures of the primary and secondary range from $\sim$ 36500--37900 K and 37500--39000 K respectively.  A non-negligible deviation from synchronous rotation was measured for the primary with an asynchronicity parameter ranging from 1.1--1.3, however the secondary did not show such a signal.  The derived helium surface abundance is lower than expected when compared to the \citet{Brott2011a} evolutionary tracks, reaching an upper limit of 0.07 while being unconstrained for the lower limit, as it reached the limits of our explored parameter range. The carbon abundance ranged from 8.7--9.0 while the nitrogen abundance ranged from 6.3--7.6.  Only a lower limit of 8.4 could be placed on the oxygen abundance, however. Calculating the luminosity as described in Equation 7, results in $\log(L_\mathrm{3D}/L_\odot) \approx $ 5.16--5.23 and 5.09--5.16 for the primary and secondary respectively.

\subsubsection{VFTS 352}
The derived temperatures for the primary and secondary range from $\sim$ 43000--45250 K and 40650--42650 K respectively.  The rotation rates of both components are compatible with tidal locking so additional broadening was not needed to reproduce the observed spectra.  To avoid degeneracies that arise from the winds, we only fit lines in the optical portion of the spectrum to derive surface chemical abundances.  The helium surface abundance for the system ranged from 0.06--0.08 while the surface abundances of carbon and nitrogen range from 6.7--7.8 and 6.2--7.8 respectively.  Since there are no oxygen lines in the wavelength range covered by the optical spectrum, the oxygen abundance is unconstrained.  The derived luminosities are $\log(L_\mathrm{3D}/L_\odot) \approx $ 5.20--5.29 and 5.11--5.19 for the primary and secondary respectively.

\subsubsection{SMC 108086}
The temperatures of the primary and secondary components of SMC 108086 range from $\sim$ 35350--36700 K and 34500--35850 K respectively.  As with V382 Cyg, the cooler component showed additional broadening with $F$ ranging from 1.13--1.35.  The helium surface abundance ranges from 0.06--0.12 and the surface abundances of carbon, nitrogen and oxygen range from 6.0--7.8, 6.0--8.1 and 6.0--8.8 respectively.  While there is structure in the oxygen abundance chi square plots, all abundances fell within the error bars so oxygen abundance remains unconstrained.   The calculated luminosities for the two components are very similar to one another with $\log(L_\mathrm{3D}/L_\odot) \approx $ 4.58--4.68 for both.

   \begin{figure*}
   \centering
   \includegraphics[width=1\linewidth]{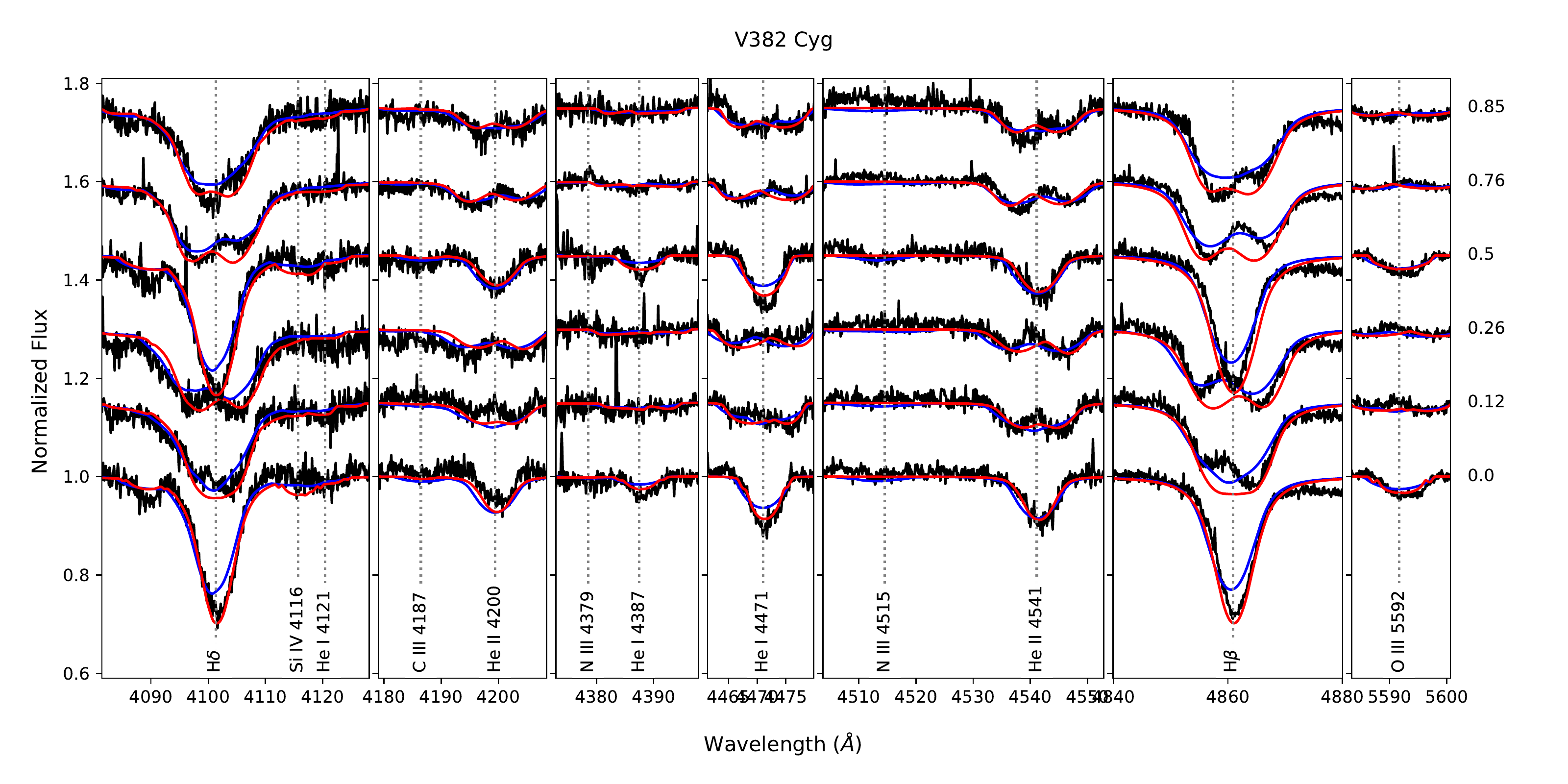}
      \caption{
      Comparison between the best fit model generated from the GA and \spamms for V382 Cyg.  The best fit solution from the GA is plotted in blue, the solution from \spamms is plotted in red and the observed spectrum is plotted in black for several orbital phases, which are indicated on the right of the plot. The locations of relevant spectral lines are indicated via vertical dashed lines.
              }
         \label{v382cyg_fit}
   \end{figure*}

   \begin{figure*}
   \centering
   \includegraphics[width=1\linewidth]{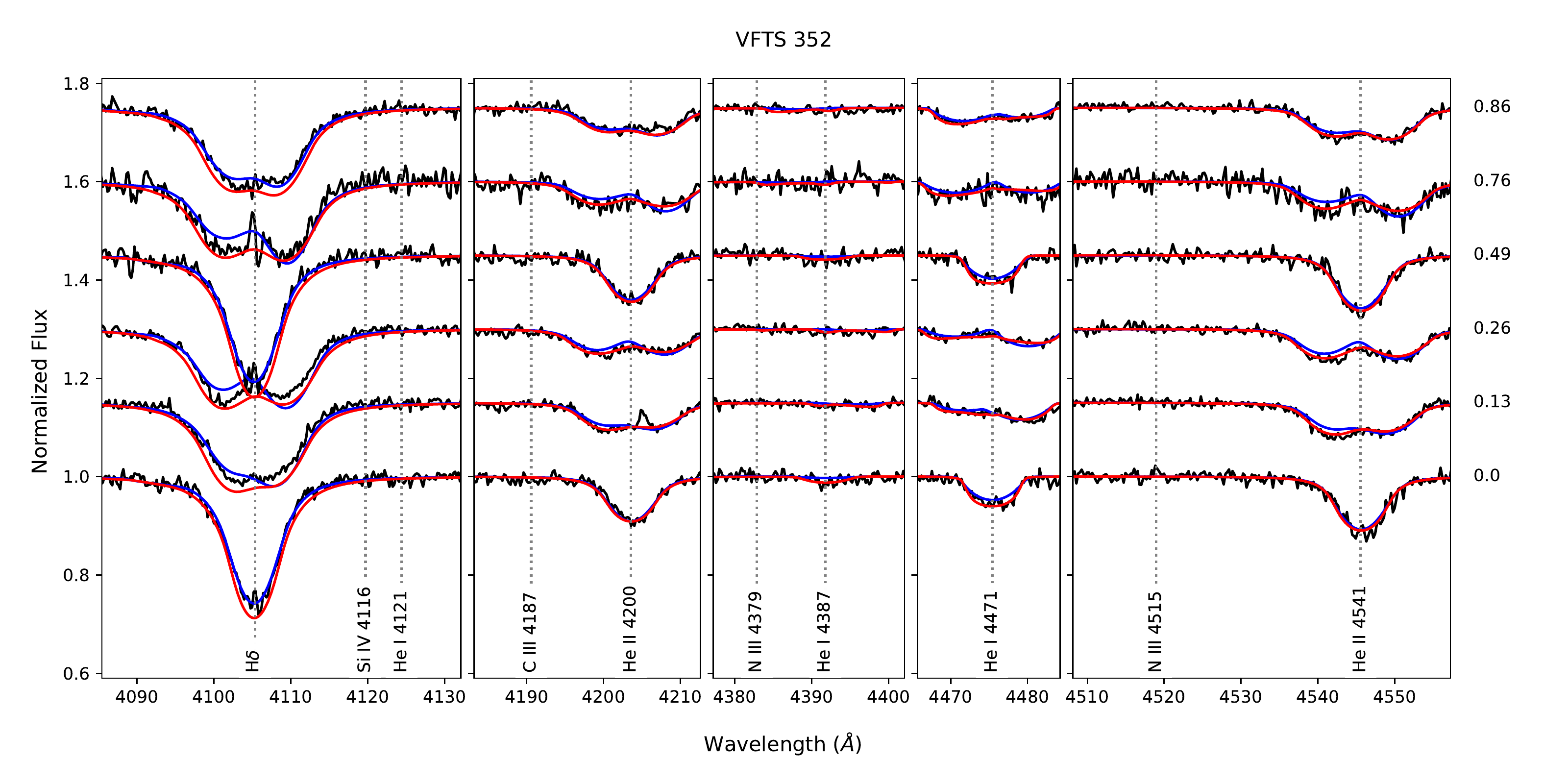}
      \caption{Same as Fig. \ref{v382cyg_fit} but for VFTS 352 instead.
              }
         \label{vfts352_fit}
   \end{figure*}

   \begin{figure*}
   \centering
   \includegraphics[width=1\linewidth]{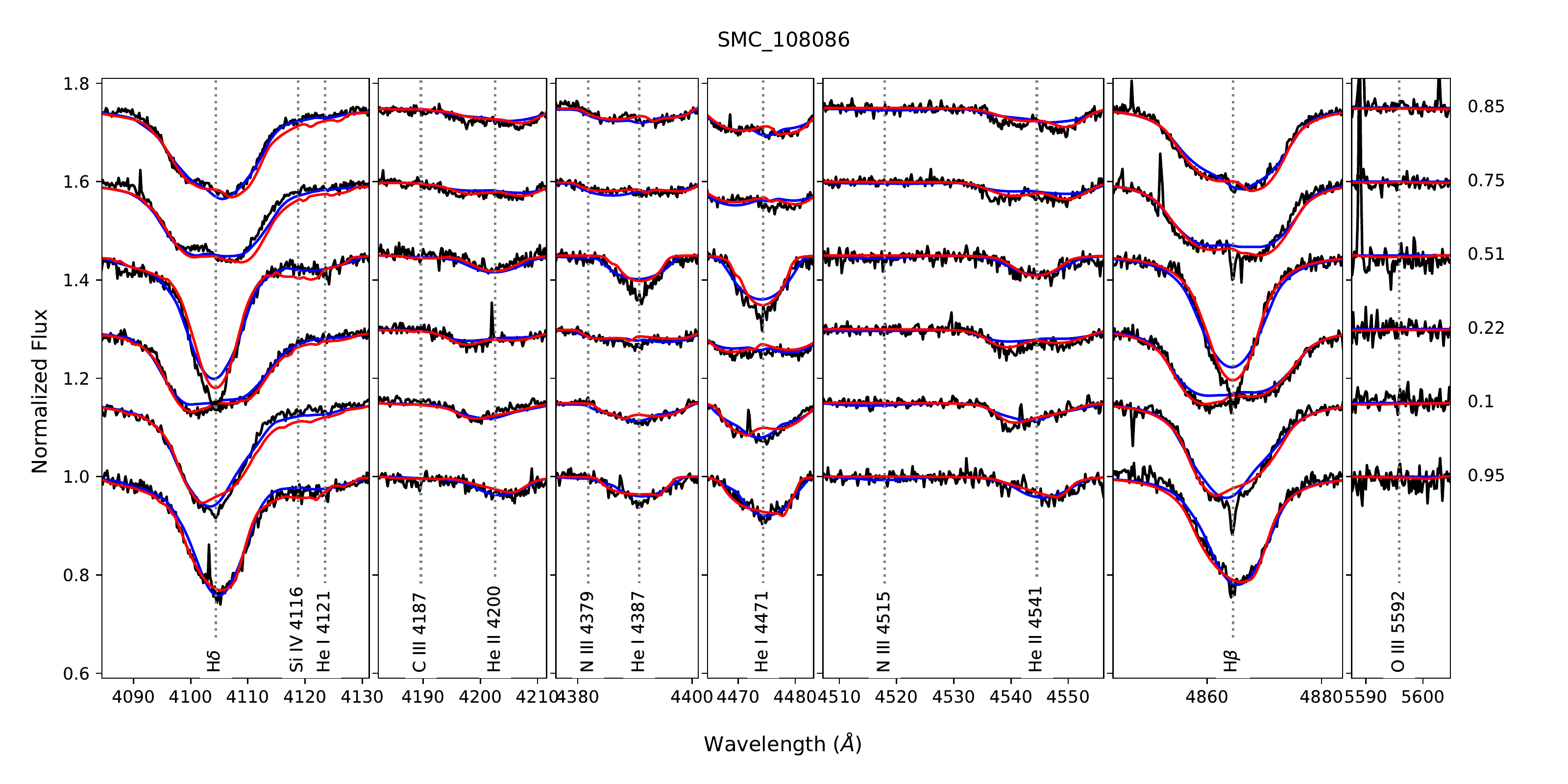}
      \caption{
      Same as Fig. \ref{v382cyg_fit} but for SMC 108086 instead.
              }
         \label{smc108086_fit}
   \end{figure*}

\section{Discussion} \label{sec:discussion}

\subsection{1D versus 3D approach}  \label{sec:discussion_3d_1d}
Both analysis methods have their own assumptions, advantages and drawbacks.  The biggest and most obvious difference between the GA and \spamms is in the assumption of the surface geometry and properties.  The 1D approach first requires spectral disentangling which not only assumes that the two stars are point sources, but also assumes that the spectral signature does not change as a function of phase (with the exception of light ratio and radial velocity variations).  This means that temperature differences across the surface are washed out and that the bridge is ignored completely.  Additionally, since the bridge is not accounted for, its signal gets shared between the two components, and this can add an artificial broadening effect to the disentangled spectra.  This effect can be seen when comparing the rotation rates measured using the two methods: the rotation rates measured with the GA a are systematically higher than those measured with \textsc{spamms}, even when the asynchronicity parameter is employed.

The spherical assumption is reinforced during the spectral fitting procedure as the \fw models assume spherical symmetry. Conversely, while the patch model still relies on 1D \fw models, it takes the surface geometry of the system into account.  This surface geometry is computed assuming Roche formalism, which has its own assumptions, an important one being that radiative accelerations are not accounted for, but it is still a significant improvement over spherical.  The fact that individual spectra are assigned to each patch across the surface means that temperature variations and the effect of the bridge are taken into account during the analysis process, although the exact gravity darkening law that should be used is still debated.  Furthermore, individual radial velocities are also assigned to each patch so there is no need to convolve the final integrated synthetic line profiles with a rotational broadening kernel as is needed when fitting using the GA.  While these formalisms result in a negligible difference in the spectral lines for spherically symmetric stars \citep[see ][]{Abdul-Masih2020a}, this will have a larger effect in contact binaries since the radial velocity of the bridge, in combination with the geometry, can be accounted for more realistically.

The GA and \spamms treat the winds in different ways. In essence, the 1D case assumes two separate wind contributions, which co-rotate and move with each star in the orbit.  The wind contributions are shifted with the same radial velocity as the corresponding star and they are rotationally broadened in the same way as the photosphere.  The patch model treats the winds in the same way that it treats the photosphere with the caveat that the mesh associated with the wind does not eclipse itself.  The validity of this assumption is discussed in detail in \citet{Abdul-Masih2020a}.  Additionally, while the patch model assumes a \citet{Vink2001} mass-loss prescription, for the 1D case, the wind parameters can be easily varied and fit.

Upon inspection of the resulting best-fit solutions from the two methods (see Figs. \ref{v382cyg_fit}, \ref{vfts352_fit}, and \ref{smc108086_fit}), it is clear that \spamms is better able to reproduce most of the observed line profiles than the GA, despite having significantly fewer degrees of freedom. When comparing the two, the line depths are better fit at 0 and 0.5 phase and the bridge (the region between the line peaks) is much better fit at quadrature with \spamms than with the GA.  Focusing on the Balmer lines, both methods struggle to reproduce the line profiles perfectly (especially in the case of V382 Cyg), however, altering the mass-loss rate in \spamms could alter the cores of the Balmer lines resulting in better agreement between the model and the observations.  Additionally, the asymmetric lines in SMC 108086 are also accurately reproduced with \spamms.

The resulting best-fit parameters derived using the two methods are more or less in agreement.  The temperatures measured using \spamms appear to be systematically higher than those derived using the GA, but the effect appears small (on the order of $\sim$ 1000 K).  As demonstrated in \citet{Abdul-Masih2020a}, this is most likely an inclination effect: when compared with 1D fitting techniques, systems with high inclination will result in higher \spamms temperatures, while systems with low inclinations will result in lower \spamms temperatures, with equivalent temperatures being reached at around an inclination of 60$^\circ$.  This effect can be seen here, where V382 Cyg and SMC 108086 with inclinations of 85$^\circ$ and 82.5$^\circ$ respectively, show higher temperatures measured using \spamms than the GA, while VFTS 352, with an inclination of 55.6$^\circ$ shows almost the exact same temperatures measured using the two techniques.

The surface abundances also match well, however \spamms does not appear to reach the same level of precision as the GA for VFTS 352. This is most likely because the GA fit of VFTS 352 included the UV spectra, however the \spamms fit only included the optical spectra.  Interestingly, however, the best fit nitrogen surface abundance derived by \spamms for VFTS 352 appears more realistic than the best fit solution from \citet{Abdul-Masih2019}, which finds a nitrogen surface abundance well below the expected baseline value for the LMC.  Similarly, in each of the three systems studied, while the ranges of the surface abundances are in good agreement, the best fit solutions within the ranges differ slightly between the two approaches.  It should be noted, however that there is a disagreement in the surface abundances of carbon for V382 Cyg: \spamms returns a higher carbon abundance than the GA by about half a dex.

While the error ranges for the helium surface abundance measured by \spamms and the GA overlap, \spamms measures very low helium surface abundances for V382 Cyg and VFTS 352.  The values returned from the GA span a wider range and appear much more realistic than the values measured by \spamms, which are below the initial Big Bang helium abundance.  One possible explanation for this arises from the fact that \spamms uses a fixed geometry, and thus a fixed surface gravity structure across the surface.  Since the surface gravity affects the ionization balance, including some of the binary parameters as free parameters, such as the component radii or the fillout factor, could raise the measured helium surface abundance and fix this issue.  This would indicate that the confidence interval returned by \spamms for the helium surface abundances are most likely underestimated.

\subsection{Evolutionary Status}
The first striking evolutionary question that arises from our spectroscopic analyses involves the fact that the spectral lines of the cooler component stars of V382 Cyg and SMC 108086 appear significantly broader than expected when assuming tidal locking.  For systems as close as these, the synchronization timescales are expected to be extremely short, so we do not expect that they are rotating asynchronously.  On the other hand, recent studies have shown that ongoing mass transfer in unequal mass ratio binaries can potentially lead to spin up for the mass gainer \citep{Menon2020}.  The fact that two of the three systems in our sample require additional broadening to properly reproduce the observed spectra implies that either these systems are still actively undergoing mass transfer, that the tidal effects behave differently in these systems or that an additional unaccounted for physical effect that mimics the effects of rotational broadening is present.

One potential explanation is macroturbulence, which we do not fit in this analysis.  While the effects of macroturbulence and rotation on spectral lines can be degenerate with each other, the macroturbulence required to explain the observed difference is quite high.  Comparing the observed broadening with the tidally locked rotation rates imply that an additional broadening source on the order of 300 \kms\ or more is needed.  Macroturbulence values for O and B type stars typically fall below $\sim$ 120 \kms\ \citep{Simon-Diaz2017}, however additional turbulent processes during the overcontact phase may justify these higher than expected values.

Since grids of evolutionary tracks suited for massive overcontact binaries are not fully developed yet, we compare our observations with single star evolutionary tracks from \citet{Brott2011a}.  While there is still debate over which mixing mechanism is dominant in massive stars \citep[see, e.g.][]{Maeder1987, Langer2012, Bowman2019}, we use the dense grid of rotation rates in the \citet{Brott2011a} tracks as a proxy for internal mixing.  Thus, higher rotation rates correspond to more internal mixing.

\subsubsection{Surface Abundances}

	\begin{figure*}
   \centering
   \includegraphics[width=1\linewidth]{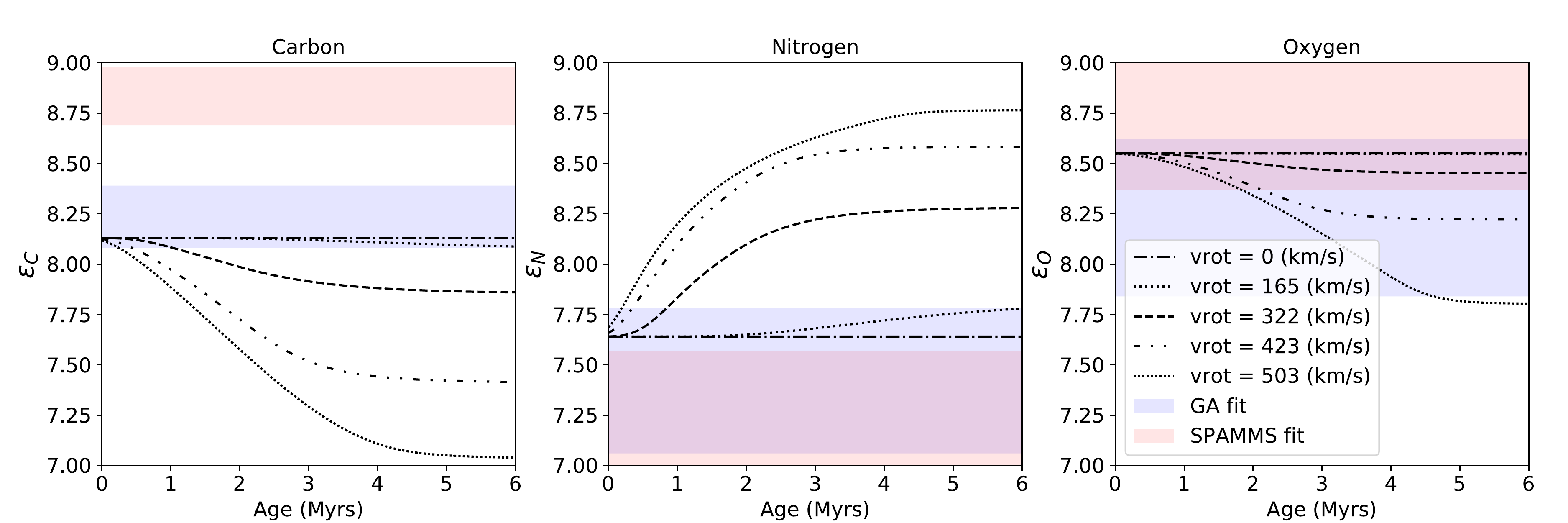}\\
   \includegraphics[width=1\linewidth]{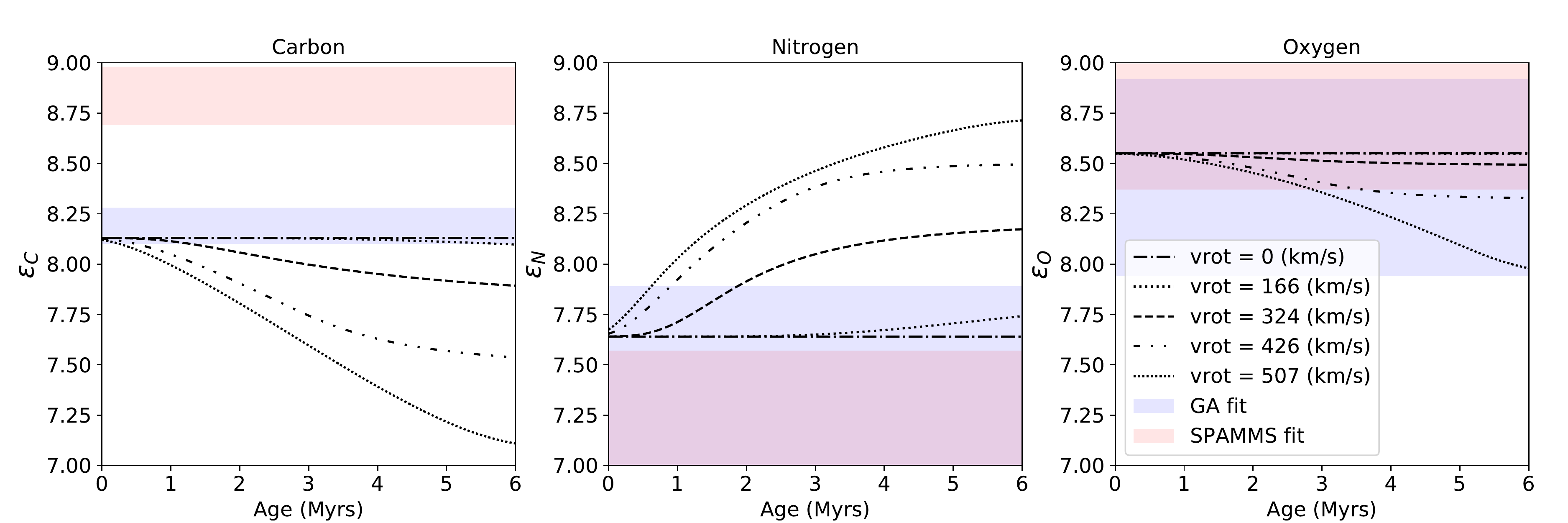}
      \caption{Surface abundances for V382 Cyg of carbon (left), nitrogen (middle), and oxygen (right), for the primary (top) and secondary star (bottom) are plotted as a function of age. Single-star Galactic-metallicity evolutionary tracks from \citet{Brott2011a} (black lines) with initial masses of $25\,\mathrm{M}_{\odot}$ for the primary and $20\,\mathrm{M}_{\odot}$ for the secondary with various initial rotational velocities are overplotted.  The blue shaded regions are the inferred range of surface abundances produced by the GA and the red shaded regions are the surface abundance ranges produced by \textsc{spamms}.  Note that the shaded region from the GA represent a 2 sigma confidence interval while the shaded region from \spamms represents a 1 sigma confidence interval.
              }
         \label{fig:abundances_v382}
   \end{figure*}
   
    \begin{figure*}
   \centering
   \includegraphics[width=1\linewidth]{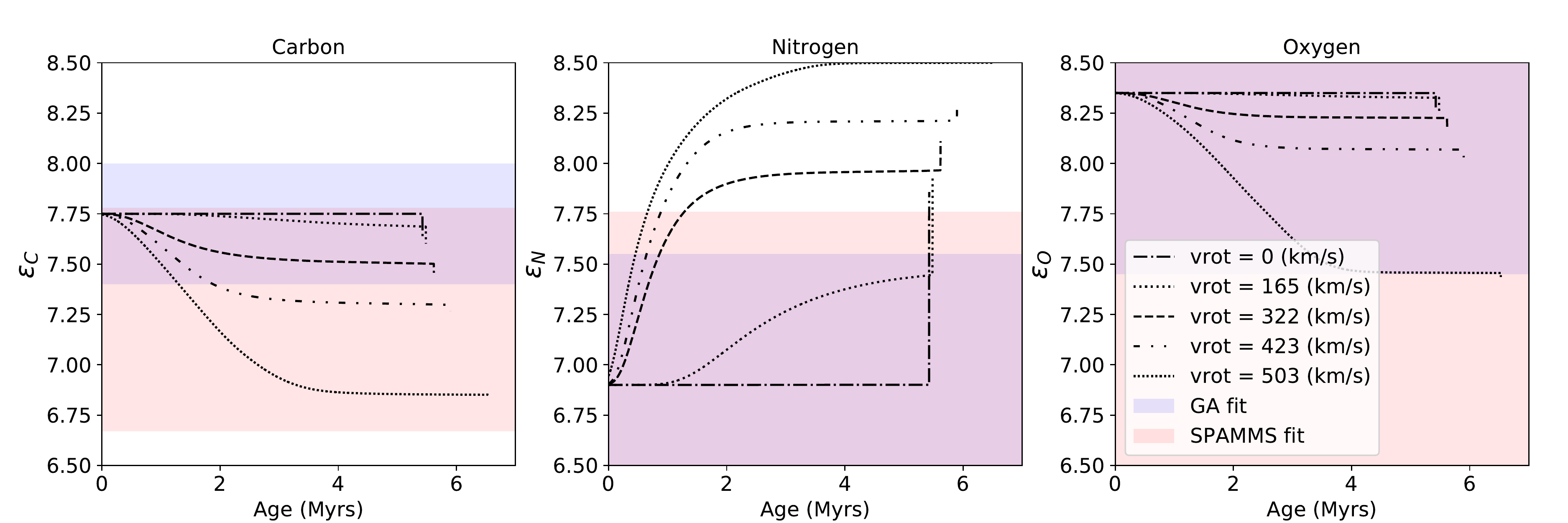}\\
   \includegraphics[width=1\linewidth]{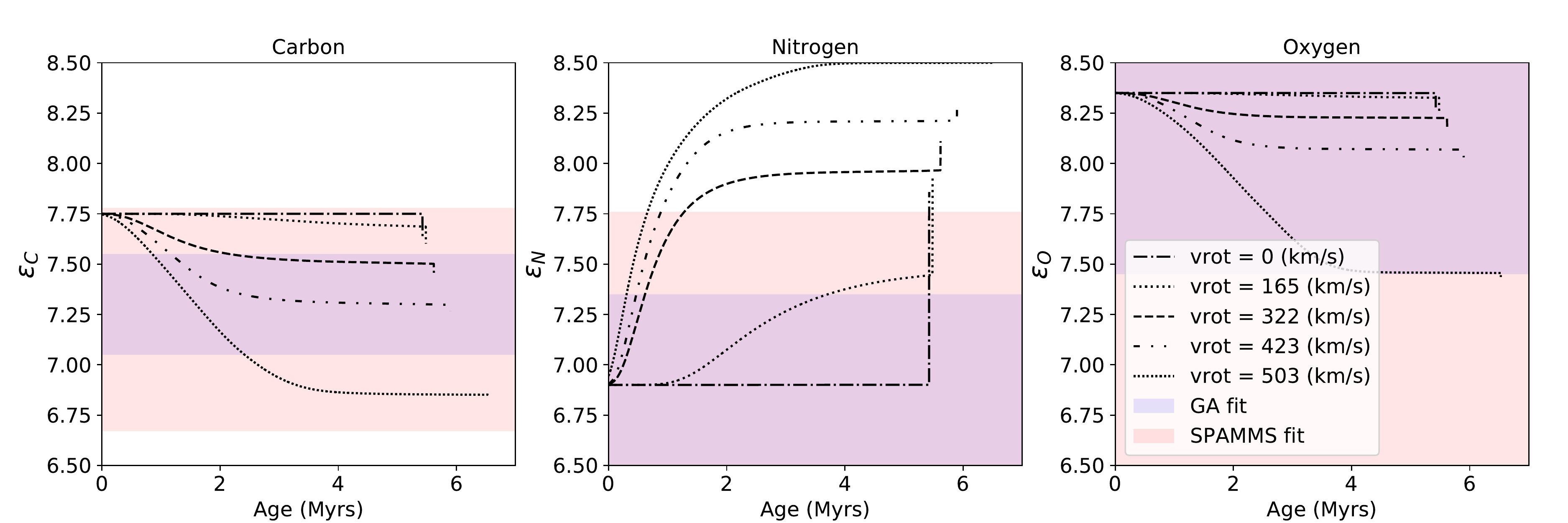}
      \caption{Same as Fig. \ref{fig:abundances_v382} but for VFTS 352.  In this case, the \citet{Brott2011a} tracks instead use LMC metallicity and initial masses of $30\,\mathrm{M}_{\odot}$ for both the primary and secondary.  Note that the GA results were taken from \citet{Abdul-Masih2019}, which used a simultaneous UV and optical fit while the \spamms fitting performed here only used the optical data.
              }
         \label{fig:abundances_vfts}
   \end{figure*}

    \begin{figure*}
   \centering
   \includegraphics[width=1\linewidth]{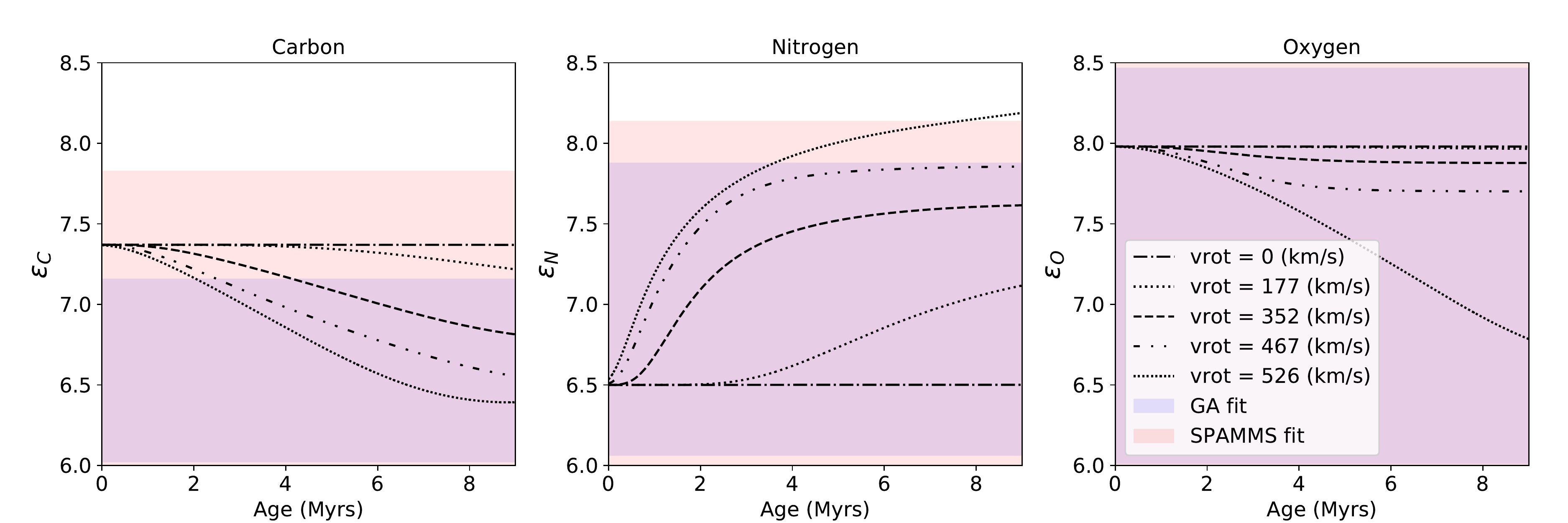}\\
   \includegraphics[width=1\linewidth]{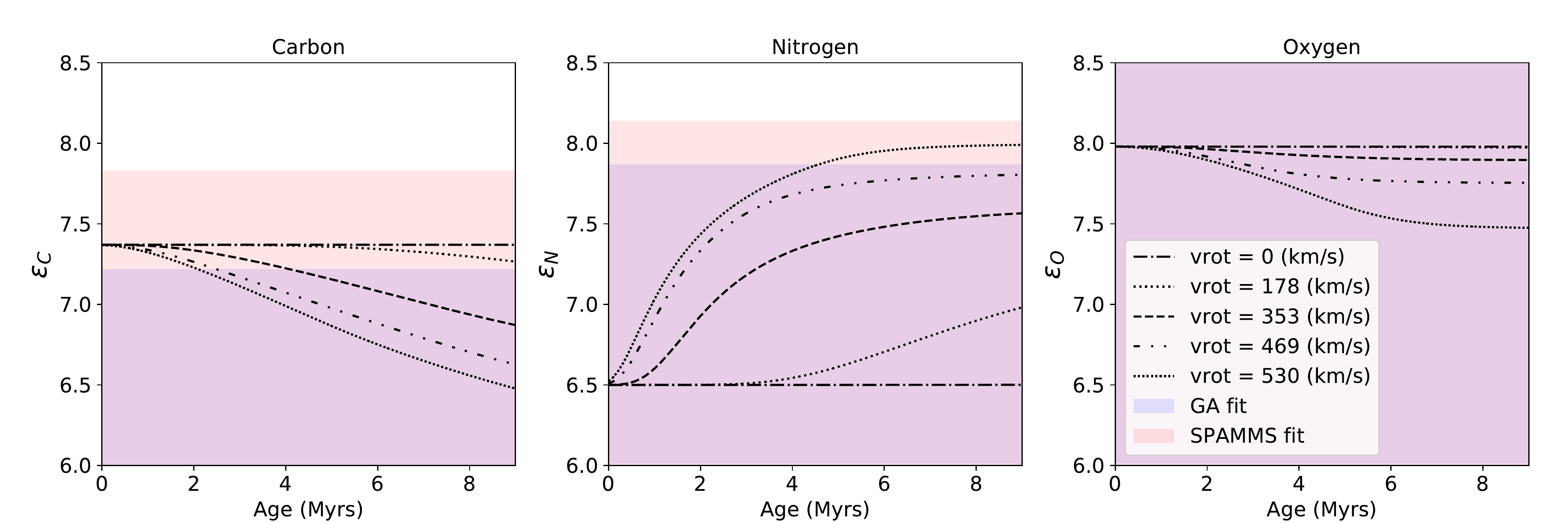}
      \caption{Same as Fig. \ref{fig:abundances_v382} but for SMC 108086.  In this case, the \citet{Brott2011a} tracks instead use SMC metallicity and initial masses of $17\,\mathrm{M}_{\odot}$ for the primary and $15\,\mathrm{M}_{\odot}$ for the secondary.
              }
         \label{fig:abundances_smc}
   \end{figure*}
   
Comparing the surface abundances measured using the GA and those measured using \textsc{spamms}, we find that the two methods are reasonably consistent.  Figures \ref{fig:abundances_v382}, \ref{fig:abundances_vfts} and \ref{fig:abundances_smc} show a comparison between the two methods for the carbon, nitrogen, and oxygen surface abundances.  Evolutionary tracks corresponding to the component masses from \citet{Brott2011a} are overplotted for various rotation rates.  The confidence intervals for the GA and \spamms are indicated with blue and red shaded regions respectively.

The upper and lower surface abundance limits for our \spamms input grid do not allow us to probe higher than 9.0 and lower than 6.0, however all of our abundances are expected to be within this range.  There are some cases where the confidence interval includes the upper or lower limit, so in these cases, the confidence interval is underestimated.  Accounting for this, the ranges obtained from the GA and \spamms are in good agreement. The biggest differences can be seen in the carbon surface abundances, which are higher when fitting with \spamms for V382 Cyg.  The opposite occurs for VFTS 352, however the line list used for the GA and for \spamms are different in this case.  Since the UV lines are affected by the winds, we do not fit these with the patch model, limiting our diagnostic lines to those in the optical portion of the spectrum.  Additionally, since the optical data for VFTS 352 only extends to $\sim$ 4550 \AA, we are not able to include several CNO lines that are included for the other two objects.  It should be noted that the increased CNO abundance grid precision for the second round of \spamms fitting did not appear to change the best fit values or the error bars in any of our runs, indicating that our initial grid precision of 0.5 dex is enough to accurately probe the parameter space using this method.

Comparing with single star evolutionary tracks, none of the systems show strong evidence of internal mixing when examining the surface abundance measurements.  V382 Cyg shows carbon, nitrogen, and oxygen all at baseline Galactic abundance without rotation (with the exception of the carbon abundance measured by \textsc{spamms}, which is above the Galactic baseline instead of below as expected from rotational mixing).  The rotation rates for the primary and secondary derived from the GA and from \spamms indicate that the two components should fall somewhere between the $\sim$ 320 \kms and the $\sim$ 420 \kms lines.  Comparing these rotation rates and the measured surface abundances with the \citet{Brott2011a} evolution tracks indicates that the age of the system must be less than one million years old.  As discussed in \citet{Abdul-Masih2019}, VFTS 352 also does not show strong signs of mixing in the CNO surface abundances either.  Only the carbon abundance indicates possible depletion, but as discussed, the surface abundance derived from \spamms was based on a smaller line list only in the optical.  The low metallicity and higher than expected rotation rates of the component stars in SMC 108086 make it difficult to constrain the surface abundances with the available data.  The results from the GA indicate a slight depletion of carbon but the confidence interval computed by \spamms extends up to baseline.  The confidence intervals for the nitrogen and oxygen surface abundances are also too wide to distinguish between the baseline and internal mixing cases for both the GA and \textsc{spamms}.

\subsubsection{Location in the HRD}

   \begin{figure*}
   \centering
   \includegraphics[width=1\linewidth]{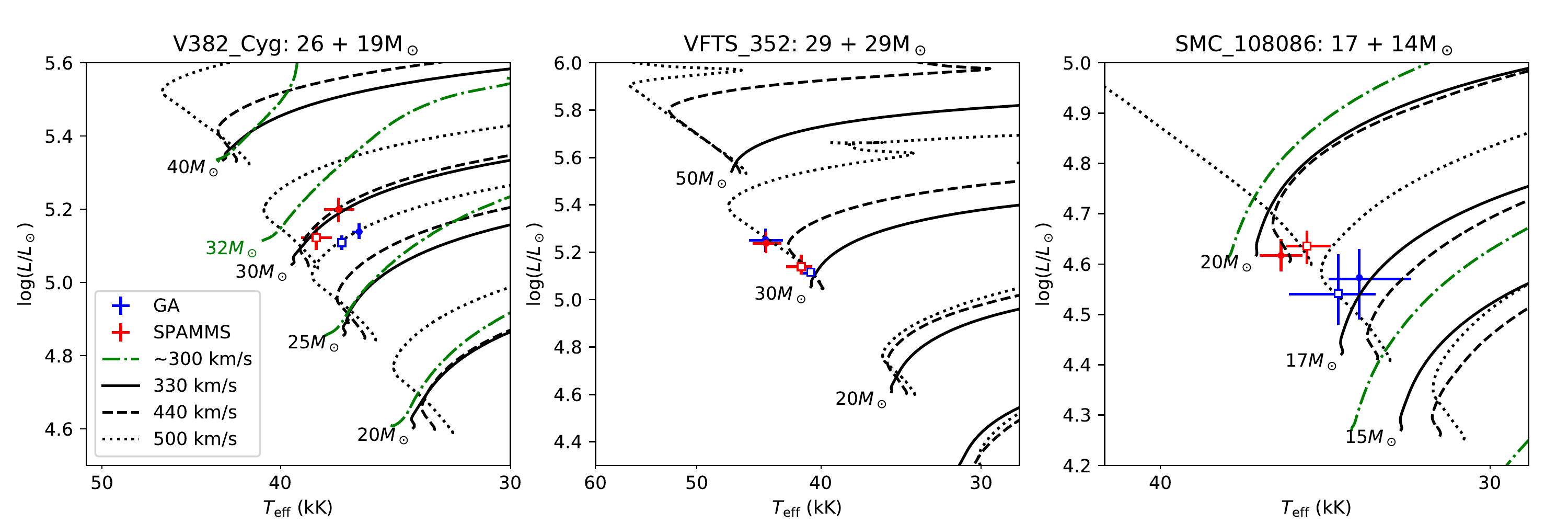}
      \caption{Locations of each component of our sample in an HRD with \citet{Brott2011a} evolutionary tracks corresponding to its metallicity.  Objects and the component dynamic masses are labeled above each panel and \citet{Brott2011a} evolutionary tracks at various rotation rates for various masses are plotted.  Values and corresponding errors derived from the GA are plotted in blue and from \spamms are plotted in red.  The primary components are indicated with filled circles, while the secondaries are indicated with open squares.  In the case of V382 Cyg and SMC 108086, rotating Geneva evolutionary tracks with initial masses of 40, 32, 25 and 20 for V382 Cyg \citep{Ekstrom2012} and 20, 15 and 12 for SMC 108086 \citep{Georgy2013} are plotted in green.
              }
         \label{hrds}
   \end{figure*}

   \begin{figure*}
   \centering
   \includegraphics[width=1\linewidth]{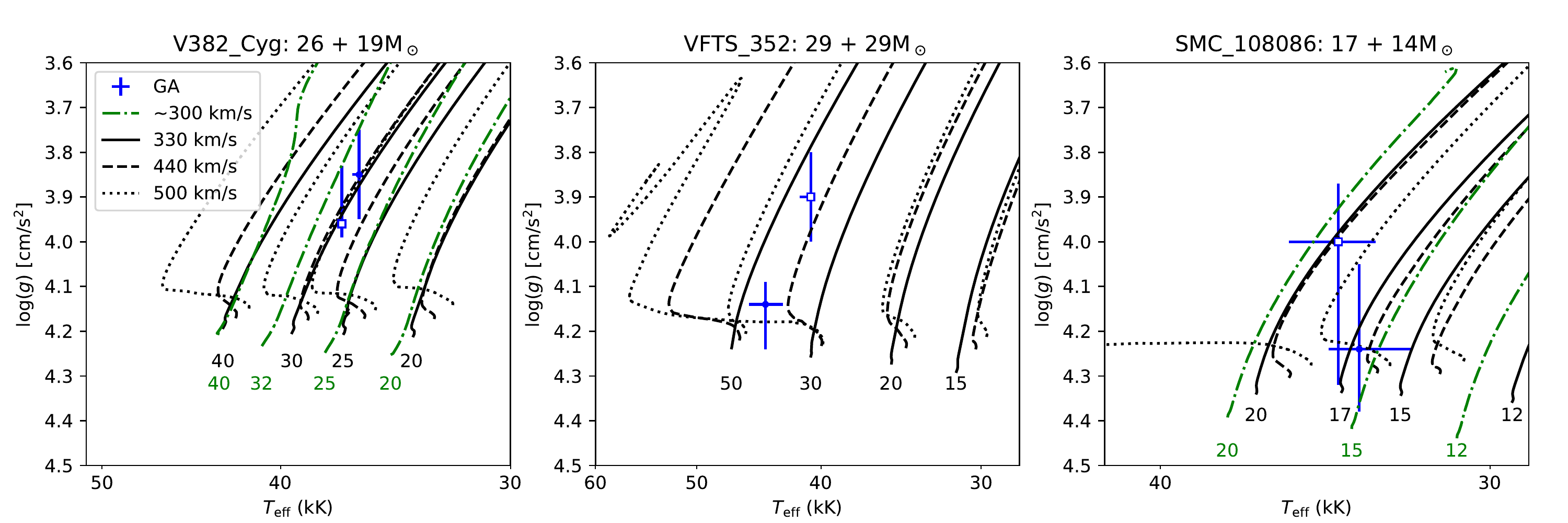}
      \caption{Same as Fig. \ref{hrds}, but in the surface gravity versus effective temperature plane instead. In this case, only the values and corresponding errors derived from the GA are plotted, since the surface gravity is not fitted with \spamms.  The masses corresponding to each track are indicated below each in their respective color in units of solar masses.
              }
         \label{spec_hrds}
   \end{figure*}
%Placing the primary and secondary components of each system in our sample on corresponding Hertzsbrung-Russel diagrams (HRDs henceforth) allows us to determine whether the systems show evidence for enhanced mixing.  According to mixing theory, as a star mixes, its surface temperature should increase.  If the mixing is efficient enough, the star may enter the qCHE regime where the star initially evolves up and to the left along the ZAMS before turning to the red and evolving more typically.  Thus, if the temperatures are higher than expected for a non-rotating model, this indicates that mixing is occurring.  We can roughly quantify this by matching the observations with rotating models to constrain the amount of internal mixing.

%Figure \ref{hrds} shows HRDs corresponding to each object in our sample.  The temperatures measured using the1D method are labeled in red and the temperatures measured using \spamms are labeled in blue.  Evolutionary tracks at various rotation rates for several masses from \citet{Brott2011a} are are plotted and the masses of the components are given above each panel for convenience.

Figure \ref{hrds} shows Hertzsprung-Russell diagrams (HRDs henceforth) corresponding to each object in our sample.  From this figure, it is clear that all three overcontact systems studied here are highly overluminous and too hot for their current masses when compared to single star evolutionary models.  Furthermore, in all three cases, the primary and secondary components fall very close to each other on the HRD.  In the case of VFTS 352, both components are of equal mass so this is expected. However for V382 Cyg and SMC 108086, the component masses are quite different with mass ratios of 1.376 and 1.183 respectively \citep{Martins2017, Hilditch2005}.  For V382 Cyg, the mass of the secondary is measured to be 19 \msol\ \citep{Martins2017}, however even with very efficient mixing, its location on the HRD cannot be reproduced with single star evolutionary models as is claimed by \citet{Martins2017}.  Instead, based on the temperature and luminosity,  the less massive component falls closer to the evolutionary tracks corresponding to the more massive component than its own.  The same thing can be seen in SMC 108086, where the secondary, with a measured mass of 14 \msol\ \citep{Hilditch2005}, falls along the 17 or 20 \msol\ tracks instead.  These mass ratios come from the dynamical masses of these systems, which are widely considered more accurate and reliable than any other mass determination techniques \citep{Serenelli2020}.  This indicates that the locations of the secondary components are indeed anomalous.  The same conclusions can be reached when comparing the locations of these objects on the $\log(g)-T_\mathrm{eff}$ plain with expectations from evolutionary tracks as shown in Fig. \ref{spec_hrds}.  

As an independent verification of the objects' locations on the HRDs, we also calculate the luminosity of the components of each system based on its V-band magnitude, distance and extinction. For V382 Cyg, we use the distance derived from GAIA early Data Release 3 \citep[$\sim$1.7 kpc][]{Bailer-Jones2021}, however since VFTS 352 and SMC 108086 are extra-Galactic systems, we instead use the distances derived from the Araucaria Project of 49.6 kpc \citep{Pietrzynski2019} and 62.1 kpc \citep{Graczyk2020} for the LMC and SMC respectively. For all three components, we use the extinction from \citet{MaizApellaniz2018}, and we compute the bolometric corrections using the relation provided by \citet{Martins2006} based on the effective temperature of the star. In this case, we use the temperatures derived from \textsc{spamms}. For V382 Cyg,  we find luminosities of $\log(L_\mathrm{dist}/L_\odot) \approx 5.09 \pm 0.05$ and $5.10 \pm 0.05$ for the primary and secondary respectively.  For VFTS 352,  we find luminosities of $\log(L_\mathrm{dist}/L_\odot) \approx 5.24 \pm 0.04$ for the primary component and $5.16 \pm 0.04$ for secondary component. Finally for SMC 108086, we calculate luminosities of $\log(L_\mathrm{dist}/L_\odot) \approx 4.60 \pm 0.03$ and $4.66 \pm 0.03$ for the primary and secondary respectively. All of these values are in very good agreement with the values we calculate using Stefan-Boltzmann with the GA results and the modified Stefan-Boltzmann using the \spamms results. 

This leads to several interesting possible implications.  Single star evolutionary models show that the less massive component should not be able to reach the temperature observed at its surface.  This means that the observed temperature of the secondary is either being driven by the flux of the primary via irradiation or heat exchange in the shared envelope, or being driven by efficient mixing in the envelope via convection or large scale circulations.  If the flux from the primary was driving the temperature of the secondary via irradiation, then we might expect to see a region of lower temperature when the non-illuminated side of the secondary is facing towards us.  This is not observed, however an indirect consequence of this mutual irradiation is a change in the net flux leaving the irradiated regions of the star.  If the fluxes leaving these regions are reduced and the overall luminosity remains constant, then the temperature of the entire star will increase slightly to compensate for this effect.  This could partially explain the increased temperature observed in the secondary star.

Another potential way to get an overluminous secondary is efficient heat exchange through the common envelope.  In the low mass counterparts of massive overcontact binaries, namely W UMa systems, overluminous secondaries are observed often.  In these systems, the overluminosity of the secondary is attributed to "sideways convection" through the shared envelope \citep{Lucy1968a, Lucy1968b}, with up to a third of the energy generated by the primary being transported to the secondary \citep{Mochnacki1981}.  Of course in the case of massive overcontact systems, the envelope is radiative instead of convective so one can imagine a similar mechanism involving heat transfer via radiation instead of convection.

Alternatively, this temperature increase could be explained by efficient internal mixing.  In rapidly rotating single stars, the difference in temperature between the pole and equator is known to drive large scale circulations in the envelope known as Eddington-Sweet circulations \citep{Eddington1925, Sweet1950}.  These occur because a rotating star cannot be in both hydrostatic and radiative thermal equilibrium at the same time \citep{vonZeipel1924}.  These Eddington-Sweet circulations operate along the temperature gradient and thus along meridional lines.  Like rapidly rotating single stars, contact systems cannot be in both hydrostatic and radiative thermal equilibrium simultaneously, and thus large scale Eddington-Sweet-like circulations should operate.  Unlike rotating single stars, however, contact systems are not azimuthally symmetric meaning that these circulations will not necessarily operate along the meridional lines.  This could allow heat exchange and material to be mixed through the bridge of the contact system and could explain the observed temperature profile.  This effect can only be properly accounted for in 3D since it arises from the breaking of both polar and azimuthal symmetry.  For a more in-depth discussion of this effect and its implications, see \citet{Hastings2020}.

While mixing, heat transfer and irradiation effects each could account for the increased temperatures and luminosities in the secondary, of the three mechanisms, only the mixing is able to account for the increased luminosity of the system as a whole.  Alternatively, this can be explained via non-conservative mass transfer as the stars initially come into contact.  If some of the envelope of the donor star was ejected from the system, this would raise the overall mean molecular weight of the system, which would in turn increase the luminosity of both components.  This would imply that the system started with a much higher total mass than is currently observed, and would suggest that some of this missing material may be observable in the surrounding environment.  It is still unclear at the current time which mechanism or combination of mechanisms is responsible for these observed effects however.

%If these circulations do not align exactly with the meridional lines, then there would be a component tangential to the surface in the same direction as the rotation.  This could explain both the observed temperature profiles and the observed asynchronous rotation rates.  This effect can only be properly accounted for in 3D since it arises from the breaking of both polar and azimuthal symmetry.

\section{Conclusions} \label{sec:summary}
We have performed a full atmospheric analysis of three massive overcontact systems using two separate analysis methods.  Using optical data sets, we have performed fits using a standard one-dimensional approach and a more sophisticated three-dimensional approach. We have compared and contrasted the results of these two methods and find that, while the 3D \spamms approach better reproduces the observed line profiles, the derived stellar parameters are still in fairly good agreement with the 1D GA approach.  We find an inclination dependent variation in the derived temperatures but we find that the surface abundances are for the most part in agreement.

Our results indicate that the temperatures of unequal mass overcontact systems do not behave as expected when compared with single star evolutionary models: the temperatures of both components are similar and appear to be driven by the higher mass component.  Single star evolutionary tracks that include rotation are not able to reproduce the location of the less massive object on the HRD.  However, when the measured temperatures of the two components are plotted on the HRD, they are both consistent with the evolutionary tracks of the more massive component when assuming high rotation rates ($\gtrsim$ 500 \kms).  Additionally we find that both components of all three systems are highly overluminous, which is a potential indicator of internal mixing or non-conservative mass loss. Conversely, the surface abundance measurements do not show definitive signs of enhancements giving further credence to the non-conservative mass loss scenario.  

When comparing our results with single star evolutionary models and ignoring binary effects such as mass exchange, the derived temperatures and luminosities seem to indicate a different level of mixing than the derived surface abundances. If the temperatures and luminosities are considered alone, the internal mixing mechanisms appear to be efficient.  However, if the surface abundances are considered alone, then there is no strong indication of mixing unless (i) the systems are very young or (ii) the component stars only began rapidly rotating after their chemical gradients were formed.  It is unlikely that all stars in our sample are young, however it is unclear why these systems would begin rotating rapidly later in their lives. Despite this,  it is still unclear how the systems can have such high temperatures but show little to no surface abundance deviations.  Binary interaction effects most likely play a large roll for these systems and could account for some if not all of the discrepancies discussed here. Alternatively there may be additional unaccounted for physical processes that are affected by the unique geometry of overcontact systems in a way we do not yet understand.  More detailed evolutionary and stellar structure models of overcontact binaries are needed to confront our measurements.  In addition, a more extensive line list and additional spectral data, especially in the UV, would allow us to better constrain the surface abundances and thus the evolutionary fate of these systems.

\begin{acknowledgements}
Based on observations obtained with the HERMES spectrograph, which is supported by the Fund for Scientific Research of Flanders (FWO), Belgium; the Research Council of KU Leuven, Belgium; the Fonds National de la Recherche Scientifique (F.R.S.-FNRS), Belgium; the Royal Observatory of Belgium; the Observatoire de Genève, Switzerland; and the Thüringer Landessternwarte Tautenburg, Germany.
This work is based on data obtained at the European Southern Observatory under program IDs.  0103.D-0237, 182.D-0222, 090.D-0323, and 092.D-0136.
We acknowledge support from the FWO-Odysseus program under project G0F8H6N.
This project has received funding from the European Research Council under European Union's Horizon 2020 research programme (grant agreement No 772225).
The computational resources and services used in this work were provided by the VSC (Flemish Supercomputer Center), funded by the Research Foundation - Flanders (FWO) and the Flemish Government.
S.d.M. A.M. and S. J. were funded in part by the European Union’s Horizon 2020 research and innovation program from the European Research Council (ERC, Grant agreement No. 715063, PI de Mink), and by the Netherlands Organization for Scientific Research (NWO) as part of the Vidi research program BinWaves (639.042.728, PI de Mink).
PM acknowledges support from the FWO junior postdoctoral fellowship No. 12ZY520N.
L. M. thanks the European Space Agency (ESA) and the Belgian Federal Science Policy Office (BELSPO) for their support in the framework of the PRODEX Programme.
\end{acknowledgements}

\bibliographystyle{aa}
\bibliography{Overcontacts}

\begin{thebibliography}{91}
\expandafter\ifx\csname natexlab\endcsname\relax\def\natexlab#1{#1}\fi

\bibitem[{{Abdul-Masih} {et~al.}(2020){Abdul-Masih}, {Sana}, {Conroy},
  {Sundqvist}, {Pr{\v{s}}a}, {Kochoska}, \& {Puls}}]{Abdul-Masih2020a}
{Abdul-Masih}, M., {Sana}, H., {Conroy}, K.~E., {et~al.} 2020, \aap, 636, A59

\bibitem[{{Abdul-Masih} {et~al.}(2019){Abdul-Masih}, {Sana}, {Sundqvist},
  {Mahy}, {Menon}, {Almeida}, {De Koter}, {de Mink}, {Justham}, {Langer},
  {Puls}, {Shenar}, \& {Tramper}}]{Abdul-Masih2019}
{Abdul-Masih}, M., {Sana}, H., {Sundqvist}, J., {et~al.} 2019, \apj, 880, 115

\bibitem[{{Almeida} {et~al.}(2015){Almeida}, {Sana}, {de Mink}, {Tramper},
  {Soszy{\'n}ski}, {Langer}, {Barb{\'a}}, {Cantiello}, {Damineli}, {de Koter},
  {Garcia}, {Gr{\"a}fener}, {Herrero}, {Howarth}, {Ma{\'\i}z Apell{\'a}niz},
  {Norman}, {Ram{\'\i}rez-Agudelo}, \& {Vink}}]{Almeida2015}
{Almeida}, L.~A., {Sana}, H., {de Mink}, S.~E., {et~al.} 2015, \apj, 812, 102

\bibitem[{{Almeida} {et~al.}(2017){Almeida}, {Sana}, {Taylor}, {Barb{\'a}},
  {Bonanos}, {Crowther}, {Damineli}, {de Koter}, {de Mink}, {Evans}, {Gieles},
  {Grin}, {H{\'e}nault-Brunet}, {Langer}, {Lennon}, {Lockwood}, {Ma{\'\i}z
  Apell{\'a}niz}, {Moffat}, {Neijssel}, {Norman}, {Ram{\'\i}rez-Agudelo},
  {Richardson}, {Schootemeijer}, {Shenar}, {Soszy{\'n}ski}, {Tramper}, \&
  {Vink}}]{Almeida2017}
{Almeida}, L.~A., {Sana}, H., {Taylor}, W., {et~al.} 2017, \aap, 598, A84

\bibitem[{{Bailer-Jones} {et~al.}(2021){Bailer-Jones}, {Rybizki}, {Fouesneau},
  {Demleitner}, \& {Andrae}}]{Bailer-Jones2021}
{Bailer-Jones}, C.~A.~L., {Rybizki}, J., {Fouesneau}, M., {Demleitner}, M., \&
  {Andrae}, R. 2021, \aj, 161, 147

\bibitem[{{Bj{\"o}rklund} {et~al.}(2021){Bj{\"o}rklund}, {Sundqvist}, {Puls},
  \& {Najarro}}]{Bjorklund2021}
{Bj{\"o}rklund}, R., {Sundqvist}, J.~O., {Puls}, J., \& {Najarro}, F. 2021,
  \aap, 648, A36

\bibitem[{{Bowman} {et~al.}(2019){Bowman}, {Burssens}, {Pedersen}, {Johnston},
  {Aerts}, {Buysschaert}, {Michielsen}, {Tkachenko}, {Rogers}, {Edelmann},
  {Ratnasingam}, {Sim{\'o}n-D{\'\i}az}, {Castro}, {Moravveji}, {Pope}, {White},
  \& {De Cat}}]{Bowman2019}
{Bowman}, D.~M., {Burssens}, S., {Pedersen}, M.~G., {et~al.} 2019, Nature
  Astronomy, 3, 760

\bibitem[{{Bresolin} {et~al.}(2008){Bresolin}, {Crowther}, \&
  {Puls}}]{Bresolin2008}
{Bresolin}, F., {Crowther}, P., \& {Puls}, J. 2008, in IAU Symposium, Vol. 250,
  Massive Stars as Cosmic Engines

\bibitem[{{Brott} {et~al.}(2011){Brott}, {de Mink}, {Cantiello}, {Langer}, {de
  Koter}, {Evans}, {Hunter}, {Trundle}, \& {Vink}}]{Brott2011a}
{Brott}, I., {de Mink}, S.~E., {Cantiello}, M., {et~al.} 2011, \aap, 530, A115

\bibitem[{{Castor} {et~al.}(1975){Castor}, {Abbott}, \& {Klein}}]{Castor1975}
{Castor}, J.~I., {Abbott}, D.~C., \& {Klein}, R.~I. 1975, \apj, 195, 157

\bibitem[{{Cester} {et~al.}(1978){Cester}, {Fedel}, {Giuricin}, {Mardirossian},
  \& {Mezzetti}}]{Cester1978}
{Cester}, B., {Fedel}, B., {Giuricin}, G., {Mardirossian}, F., \& {Mezzetti},
  M. 1978, \aaps, 33, 91

\bibitem[{{Charbonneau}(1995)}]{Charbonneau1995}
{Charbonneau}, P. 1995, \apjs, 101, 309

\bibitem[{{Conroy} {et~al.}(2020){Conroy}, {Kochoska}, {Hey}, {Pablo},
  {Hambleton}, {Jones}, {Giammarco}, {Abdul-Masih}, \&
  {Pr{\v{s}}a}}]{Conroy2020}
{Conroy}, K.~E., {Kochoska}, A., {Hey}, D., {et~al.} 2020, \apjs, 250, 34

\bibitem[{Darwin(1859)}]{Darwin1859}
Darwin, C. 1859, On the Origin of Species by Means of Natural Selection
  (London: Murray), or the Preservation of Favored Races in the Struggle for
  Life

\bibitem[{{de Mink} {et~al.}(2009){de Mink}, {Cantiello}, {Langer}, {Pols},
  {Brott}, \& {Yoon}}]{deMink2009}
{de Mink}, S.~E., {Cantiello}, M., {Langer}, N., {et~al.} 2009, \aap, 497, 243

\bibitem[{{de Mink} \& {Mandel}(2016)}]{deMink2016}
{de Mink}, S.~E. \& {Mandel}, I. 2016, \mnras, 460, 3545

\bibitem[{{de Mink} {et~al.}(2007){de Mink}, {Pols}, \&
  {Hilditch}}]{deMink2007}
{de Mink}, S.~E., {Pols}, O.~R., \& {Hilditch}, R.~W. 2007, \aap, 467, 1181

\bibitem[{{De{\v{g}}irmenci} {et~al.}(1999){De{\v{g}}irmenci}, {Sezer},
  {Demircan}, {Erdem}, {{\"O}zdemir}, {Ak}, \& {Albayrak}}]{Degirmenci1999}
{De{\v{g}}irmenci}, {\"O}.~L., {Sezer}, C., {Demircan}, O., {et~al.} 1999,
  \aaps, 134, 327

\bibitem[{{du Buisson} {et~al.}(2020){du Buisson}, {Marchant}, {Podsiadlowski},
  {Kobayashi}, {Abdalla}, {Taylor}, {Mandel}, {de Mink}, {Moriya}, \&
  {Langer}}]{duBuisson2020}
{du Buisson}, L., {Marchant}, P., {Podsiadlowski}, P., {et~al.} 2020, \mnras,
  499, 5941

\bibitem[{{Eddington}(1925)}]{Eddington1925}
{Eddington}, A.~S. 1925, The Observatory, 48, 73

\bibitem[{{Eggen} \& {Iben}(1989)}]{Eggen1989}
{Eggen}, O.~J. \& {Iben}, Icko, J. 1989, \aj, 97, 431

\bibitem[{{Ekstr{\"o}m} {et~al.}(2012){Ekstr{\"o}m}, {Georgy}, {Eggenberger},
  {Meynet}, {Mowlavi}, {Wyttenbach}, {Granada}, {Decressin}, {Hirschi},
  {Frischknecht}, {Charbonnel}, \& {Maeder}}]{Ekstrom2012}
{Ekstr{\"o}m}, S., {Georgy}, C., {Eggenberger}, P., {et~al.} 2012, \aap, 537,
  A146

\bibitem[{{Freudling} {et~al.}(2013){Freudling}, {Romaniello}, {Bramich},
  {Ballester}, {Forchi}, {Garc{\'\i}a-Dabl{\'o}}, {Moehler}, \&
  {Neeser}}]{Freudling2013}
{Freudling}, W., {Romaniello}, M., {Bramich}, D.~M., {et~al.} 2013, \aap, 559,
  A96

\bibitem[{{Georgy} {et~al.}(2013){Georgy}, {Ekstr{\"o}m}, {Eggenberger},
  {Meynet}, {Haemmerl{\'e}}, {Maeder}, {Granada}, {Groh}, {Hirschi}, {Mowlavi},
  {Yusof}, {Charbonnel}, {Decressin}, \& {Barblan}}]{Georgy2013}
{Georgy}, C., {Ekstr{\"o}m}, S., {Eggenberger}, P., {et~al.} 2013, \aap, 558,
  A103

\bibitem[{{Graczyk} {et~al.}(2020){Graczyk}, {Pietrzy{\'n}ski}, {Thompson},
  {Gieren}, {Zgirski}, {Villanova}, {G{\'o}rski}, {Wielg{\'o}rski},
  {Karczmarek}, {Narloch}, {Pilecki}, {Taormina}, {Smolec}, {Suchomska},
  {Gallenne}, {Nardetto}, {Storm}, {Kudritzki}, {Ka{\l}uszy{\'n}ski}, \&
  {Pych}}]{Graczyk2020}
{Graczyk}, D., {Pietrzy{\'n}ski}, G., {Thompson}, I.~B., {et~al.} 2020, \apj,
  904, 13

\bibitem[{{Hadrava}(1995)}]{Hadrava1995}
{Hadrava}, P. 1995, \aaps, 114, 393

\bibitem[{{Hadrava}(2009)}]{Hadrava2009}
{Hadrava}, P. 2009, arXiv e-prints, arXiv:0909.0172

\bibitem[{{Hastings} {et~al.}(2020){Hastings}, {Langer}, \&
  {Koenigsberger}}]{Hastings2020}
{Hastings}, B., {Langer}, N., \& {Koenigsberger}, G. 2020, \aap, 641, A86

\bibitem[{{Hilditch} {et~al.}(2005){Hilditch}, {Howarth}, \&
  {Harries}}]{Hilditch2005}
{Hilditch}, R.~W., {Howarth}, I.~D., \& {Harries}, T.~J. 2005, \mnras, 357, 304

\bibitem[{{Horvat} {et~al.}(2018){Horvat}, {Conroy}, {Pablo}, {Hambleton},
  {Kochoska}, {Giammarco}, \& {Pr{\v{s}}a}}]{Horvat2018}
{Horvat}, M., {Conroy}, K.~E., {Pablo}, H., {et~al.} 2018, \apjs, 237, 26

\bibitem[{{Howarth} {et~al.}(2015){Howarth}, {Dufton}, {Dunstall}, {Evans},
  {Almeida}, {Bonanos}, {Clark}, {Langer}, {Sana}, {Sim{\'o}n-D{\'\i}az},
  {Soszy{\'n}ski}, \& {Taylor}}]{Howarth2015}
{Howarth}, I.~D., {Dufton}, P.~L., {Dunstall}, P.~R., {et~al.} 2015, \aap, 582,
  A73

\bibitem[{{Ilijic} {et~al.}(2004){Ilijic}, {Hensberge}, {Pavlovski}, \&
  {Freyhammer}}]{Ilijic2004}
{Ilijic}, S., {Hensberge}, H., {Pavlovski}, K., \& {Freyhammer}, L.~M. 2004, in
  Astronomical Society of the Pacific Conference Series, Vol. 318,
  Spectroscopically and Spatially Resolving the Components of the Close Binary
  Stars, ed. R.~W. {Hilditch}, H.~{Hensberge}, \& K.~{Pavlovski}, 111--113

\bibitem[{{Janssens} {et~al.}(2021){Janssens}, {Shenar}, {Mahy}, {Marchant},
  {Sana}, \& {Bodensteiner}}]{Janssens2021}
{Janssens}, S., {Shenar}, T., {Mahy}, L., {et~al.} 2021, \aap, 646, A33

\bibitem[{{Jones} {et~al.}(2020){Jones}, {Conroy}, {Horvat}, {Giammarco},
  {Kochoska}, {Pablo}, {Brown}, {Sowicka}, \& {Pr{\v{s}}a}}]{Jones2020}
{Jones}, D., {Conroy}, K.~E., {Horvat}, M., {et~al.} 2020, \apjs, 247, 63

\bibitem[{{Justham} {et~al.}(2014){Justham}, {Podsiadlowski}, \&
  {Vink}}]{Justham2014}
{Justham}, S., {Podsiadlowski}, P., \& {Vink}, J.~S. 2014, \apj, 796, 121

\bibitem[{{Langer}(2012)}]{Langer2012}
{Langer}, N. 2012, \araa, 50, 107

\bibitem[{{Leung} \& {Schneider}(1978)}]{Leung1978}
{Leung}, K.~C. \& {Schneider}, D.~P. 1978, \apj, 222, 924

\bibitem[{{Lorenzo} {et~al.}(2014){Lorenzo}, {Negueruela}, {Baker},
  {Garc{\'\i}a}, {Sim{\'o}n-D{\'\i}az}, {Pastor}, \& {M{\'e}ndez
  Majuelos}}]{Lorenzo2014}
{Lorenzo}, J., {Negueruela}, I., {Baker}, A.~K.~F.~V., {et~al.} 2014, \aap,
  572, A110

\bibitem[{{Lorenzo} {et~al.}(2017){Lorenzo}, {Sim{\'o}n-D{\'\i}az},
  {Negueruela}, {Vilardell}, {Garcia}, {Evans}, \& {Montes}}]{Lorenzo2017}
{Lorenzo}, J., {Sim{\'o}n-D{\'\i}az}, S., {Negueruela}, I., {et~al.} 2017,
  \aap, 606, A54

\bibitem[{{Lucy}(1968{\natexlab{a}})}]{Lucy1968a}
{Lucy}, L.~B. 1968{\natexlab{a}}, \apj, 153, 877

\bibitem[{{Lucy}(1968{\natexlab{b}})}]{Lucy1968b}
{Lucy}, L.~B. 1968{\natexlab{b}}, \apj, 151, 1123

\bibitem[{{Maeder}(1987)}]{Maeder1987}
{Maeder}, A. 1987, \aap, 178, 159

\bibitem[{{Mahy} {et~al.}(2020{\natexlab{a}}){Mahy}, {Almeida}, {Sana},
  {Clark}, {de Koter}, {de Mink}, {Evans}, {Grin}, {Langer}, {Moffat},
  {Schneider}, {Shenar}, \& {Tramper}}]{Mahy2020b}
{Mahy}, L., {Almeida}, L.~A., {Sana}, H., {et~al.} 2020{\natexlab{a}}, \aap,
  634, A119

\bibitem[{{Mahy} {et~al.}(2020{\natexlab{b}}){Mahy}, {Sana}, {Abdul-Masih},
  {Almeida}, {Langer}, {Shenar}, {de Koter}, {de Mink}, {de Wit}, {Grin},
  {Evans}, {Moffat}, {Schneider}, {Barb{\'a}}, {Clark}, {Crowther},
  {Gr{\"a}fener}, {Lennon}, {Tramper}, \& {Vink}}]{Mahy2020a}
{Mahy}, L., {Sana}, H., {Abdul-Masih}, M., {et~al.} 2020{\natexlab{b}}, \aap,
  634, A118

\bibitem[{{Ma{\'\i}z Apell{\'a}niz} \& {Barb{\'a}}(2018)}]{MaizApellaniz2018}
{Ma{\'\i}z Apell{\'a}niz}, J. \& {Barb{\'a}}, R.~H. 2018, \aap, 613, A9

\bibitem[{{Mandel} \& {de Mink}(2016)}]{Mandel2016}
{Mandel}, I. \& {de Mink}, S.~E. 2016, \mnras, 458, 2634

\bibitem[{{Marchant} {et~al.}(2016){Marchant}, {Langer}, {Podsiadlowski},
  {Tauris}, \& {Moriya}}]{Marchant2016}
{Marchant}, P., {Langer}, N., {Podsiadlowski}, P., {Tauris}, T.~M., \&
  {Moriya}, T.~J. 2016, \aap, 588, A50

\bibitem[{{Martins} {et~al.}(2017){Martins}, {Mahy}, \&
  {Herv{\'e}}}]{Martins2017}
{Martins}, F., {Mahy}, L., \& {Herv{\'e}}, A. 2017, \aap, 607, A82

\bibitem[{{Martins} \& {Plez}(2006)}]{Martins2006}
{Martins}, F. \& {Plez}, B. 2006, \aap, 457, 637

\bibitem[{{Mateo} {et~al.}(1990){Mateo}, {Harris}, {Nemec}, \&
  {Olszewski}}]{Mateo1990}
{Mateo}, M., {Harris}, H.~C., {Nemec}, J., \& {Olszewski}, E.~W. 1990, \aj,
  100, 469

\bibitem[{{Menon} {et~al.}(2020){Menon}, {Krishnaraj}, {Anabha}, {Devaky}, \&
  {Thomas}}]{Menon2020}
{Menon}, P.~K., {Krishnaraj}, K.~U., {Anabha}, E.~R., {Devaky}, K.~S., \&
  {Thomas}, S.~P. 2020, Journal of Molecular Structure, 1222, 128798

\bibitem[{{Mochnacki}(1981)}]{Mochnacki1981}
{Mochnacki}, S.~W. 1981, \apj, 245, 650

\bibitem[{{Mochnacki} \& {Doughty}(1972)}]{Mochnacki1972}
{Mochnacki}, S.~W. \& {Doughty}, N.~A. 1972, \mnras, 156, 51

\bibitem[{{Mokiem} {et~al.}(2007){Mokiem}, {de Koter}, {Evans}, {Puls},
  {Smartt}, {Crowther}, {Herrero}, {Langer}, {Lennon}, {Najarro}, {Villamariz},
  \& {Vink}}]{Mokiem2007}
{Mokiem}, M.~R., {de Koter}, A., {Evans}, C.~J., {et~al.} 2007, \aap, 465, 1003

\bibitem[{{Mokiem} {et~al.}(2006){Mokiem}, {de Koter}, {Evans}, {Puls},
  {Smartt}, {Crowther}, {Herrero}, {Langer}, {Lennon}, {Najarro}, {Villamariz},
  \& {Yoon}}]{Mokiem2006}
{Mokiem}, M.~R., {de Koter}, A., {Evans}, C.~J., {et~al.} 2006, \aap, 456, 1131

\bibitem[{{Mokiem} {et~al.}(2005){Mokiem}, {de Koter}, {Puls}, {Herrero},
  {Najarro}, \& {Villamariz}}]{Mokiem2005}
{Mokiem}, M.~R., {de Koter}, A., {Puls}, J., {et~al.} 2005, \aap, 441, 711

\bibitem[{{Pavlovski} \& {Hensberge}(2010)}]{Pavlovski2010}
{Pavlovski}, K. \& {Hensberge}, H. 2010, in Astronomical Society of the Pacific
  Conference Series, Vol. 435, Binaries - Key to Comprehension of the Universe,
  ed. A.~{Pr{\v{s}}a} \& M.~{Zejda}, 207

\bibitem[{{Pawlak} {et~al.}(2016){Pawlak}, {Soszy{\'n}ski}, {Udalski},
  {Szyma{\'n}ski}, {Wyrzykowski}, {Ulaczyk}, {Poleski}, {Pietrukowicz},
  {Koz{\l}owski}, {Skowron}, {Skowron}, {Mr{\'o}z}, \&
  {Hamanowicz}}]{Pawlak2016}
{Pawlak}, M., {Soszy{\'n}ski}, I., {Udalski}, A., {et~al.} 2016, \actaa, 66,
  421

\bibitem[{{Penny} {et~al.}(2008){Penny}, {Ouzts}, \& {Gies}}]{Penny2008}
{Penny}, L.~R., {Ouzts}, C., \& {Gies}, D.~R. 2008, \apj, 681, 554

\bibitem[{{Pietrzy{\'n}ski} {et~al.}(2019){Pietrzy{\'n}ski}, {Graczyk},
  {Gallenne}, {Gieren}, {Thompson}, {Pilecki}, {Karczmarek}, {G{\'o}rski},
  {Suchomska}, {Taormina}, {Zgirski}, {Wielg{\'o}rski}, {Ko{\l}aczkowski},
  {Konorski}, {Villanova}, {Nardetto}, {Kervella}, {Bresolin}, {Kudritzki},
  {Storm}, {Smolec}, \& {Narloch}}]{Pietrzynski2019}
{Pietrzy{\'n}ski}, G., {Graczyk}, D., {Gallenne}, A., {et~al.} 2019, \nat, 567,
  200

\bibitem[{{Pols}(1994)}]{Pols1994}
{Pols}, O.~R. 1994, \aap, 290, 119

\bibitem[{{Popper}(1978)}]{Popper1978}
{Popper}, D.~M. 1978, \apjl, 220, L11

\bibitem[{{Pr{\v{s}}a} {et~al.}(2016){Pr{\v{s}}a}, {Conroy}, {Horvat}, {Pablo},
  {Kochoska}, {Bloemen}, {Giammarco}, {Hambleton}, \& {Degroote}}]{Prsa2016}
{Pr{\v{s}}a}, A., {Conroy}, K.~E., {Horvat}, M., {et~al.} 2016, \apjs, 227, 29

\bibitem[{{Puls} {et~al.}(2005){Puls}, {Urbaneja}, {Venero}, {Repolust},
  {Springmann}, {Jokuthy}, \& {Mokiem}}]{Puls2005}
{Puls}, J., {Urbaneja}, M.~A., {Venero}, R., {et~al.} 2005, \aap, 435, 669

\bibitem[{{Puls} {et~al.}(2008){Puls}, {Vink}, \& {Najarro}}]{Puls2008}
{Puls}, J., {Vink}, J.~S., \& {Najarro}, F. 2008, \aapr, 16, 209

\bibitem[{{Ram{\'\i}rez-Agudelo} {et~al.}(2017){Ram{\'\i}rez-Agudelo}, {Sana},
  {de Koter}, {Tramper}, {Grin}, {Schneider}, {Langer}, {Puls}, {Markova},
  {Bestenlehner}, {Castro}, {Crowther}, {Evans}, {Garc{\'\i}a}, {Gr{\"a}fener},
  {Herrero}, {van Kempen}, {Lennon}, {Ma{\'\i}z Apell{\'a}niz}, {Najarro},
  {Sab{\'\i}n-Sanjuli{\'a}n}, {Sim{\'o}n-D{\'\i}az}, {Taylor}, \&
  {Vink}}]{Ramirez-Agudelo2017}
{Ram{\'\i}rez-Agudelo}, O.~H., {Sana}, H., {de Koter}, A., {et~al.} 2017, \aap,
  600, A81

\bibitem[{{Raskin} {et~al.}(2011){Raskin}, {van Winckel}, {Hensberge},
  {Jorissen}, {Lehmann}, {Waelkens}, {Avila}, {de Cuyper}, {Degroote},
  {Dubosson}, {Dumortier}, {Fr{\'e}mat}, {Laux}, {Michaud}, {Morren}, {Perez
  Padilla}, {Pessemier}, {Prins}, {Smolders}, {van Eck}, \&
  {Winkler}}]{Raskin2011}
{Raskin}, G., {van Winckel}, H., {Hensberge}, H., {et~al.} 2011, \aap, 526, A69

\bibitem[{{Repolust} {et~al.}(2004){Repolust}, {Puls}, \&
  {Herrero}}]{Repolust2004}
{Repolust}, T., {Puls}, J., \& {Herrero}, A. 2004, \aap, 415, 349

\bibitem[{{Riley} {et~al.}(2020){Riley}, {Mandel}, {Marchant}, {Butler},
  {Nathaniel}, {Neijssel}, {Shortt}, \& {Vigna-Gomez}}]{Riley2020}
{Riley}, J., {Mandel}, I., {Marchant}, P., {et~al.} 2020, arXiv e-prints,
  arXiv:2010.00002

\bibitem[{{Sana} {et~al.}(2012){Sana}, {de Mink}, {de Koter}, {Langer},
  {Evans}, {Gieles}, {Gosset}, {Izzard}, {Le Bouquin}, \&
  {Schneider}}]{Sana2012}
{Sana}, H., {de Mink}, S.~E., {de Koter}, A., {et~al.} 2012, Science, 337, 444

\bibitem[{{Sana} \& {Evans}(2011)}]{Sana2011}
{Sana}, H. \& {Evans}, C.~J. 2011, in Active OB Stars: Structure, Evolution,
  Mass Loss, and Critical Limits, ed. C.~{Neiner}, G.~{Wade}, G.~{Meynet}, \&
  G.~{Peters}, Vol. 272, 474--485

\bibitem[{{Schneider} {et~al.}(2019){Schneider}, {Ohlmann}, {Podsiadlowski},
  {R{\"o}pke}, {Balbus}, {Pakmor}, \& {Springel}}]{Schneider2019}
{Schneider}, F. R.~N., {Ohlmann}, S.~T., {Podsiadlowski}, P., {et~al.} 2019,
  \nat, 574, 211

\bibitem[{{Sekaran} {et~al.}(2020){Sekaran}, {Tkachenko}, {Abdul-Masih},
  {Pr{\v{s}}a}, {Johnston}, {Huber}, {Murphy}, {Banyard}, {Howard}, {Isaacson},
  {Bowman}, \& {Aerts}}]{Sekaran2020}
{Sekaran}, S., {Tkachenko}, A., {Abdul-Masih}, M., {et~al.} 2020, \aap, 643,
  A162

\bibitem[{{Serenelli} {et~al.}(2020){Serenelli}, {Weiss}, {Aerts}, {Angelou},
  {Baroch}, {Bastian}, {Beck}, {Bergemann}, {Bestenlehner}, {Czekala},
  {Elias-Rosa}, {Escorza}, {Van Eylen}, {Feuillet}, {Gandolfi}, {Gieles},
  {Girardi}, {Lebreton}, {Lodieu}, {Martig}, {Miller Bertolami}, {Mombarg},
  {Morales}, {Moya}, {Nsamba}, {Pavlovski}, {Pedersen}, {Ribas}, {Schneider},
  {Silva Aguirre}, {Stassun}, {Tolstoy}, {Tremblay}, \&
  {Zwintz}}]{Serenelli2020}
{Serenelli}, A., {Weiss}, A., {Aerts}, C., {et~al.} 2020, arXiv e-prints,
  arXiv:2006.10868

\bibitem[{{Shao} \& {Li}(2014)}]{Shao2014}
{Shao}, Y. \& {Li}, X.-D. 2014, \apj, 796, 37

\bibitem[{{Simon} \& {Sturm}(1994)}]{Simon1994}
{Simon}, K.~P. \& {Sturm}, E. 1994, \aap, 281, 286

\bibitem[{{Sim{\'o}n-D{\'\i}az} {et~al.}(2017){Sim{\'o}n-D{\'\i}az}, {Godart},
  {Castro}, {Herrero}, {Aerts}, {Puls}, {Telting}, \&
  {Grassitelli}}]{Simon-Diaz2017}
{Sim{\'o}n-D{\'\i}az}, S., {Godart}, M., {Castro}, N., {et~al.} 2017, \aap,
  597, A22

\bibitem[{{Smith} {et~al.}(2018){Smith}, {Andrews}, {Rest}, {Bianco}, {Prieto},
  {Matheson}, {James}, {Smith}, {Strampelli}, \& {Zenteno}}]{Smith2018}
{Smith}, N., {Andrews}, J.~E., {Rest}, A., {et~al.} 2018, \mnras, 480, 1466

\bibitem[{{Sundqvist} \& {Puls}(2018)}]{Sundqvist2018a}
{Sundqvist}, J.~O. \& {Puls}, J. 2018, \aap, 619, A59

\bibitem[{{Sweet}(1950)}]{Sweet1950}
{Sweet}, P.~A. 1950, \mnras, 110, 548

\bibitem[{{Sz{\'e}csi} {et~al.}(2015){Sz{\'e}csi}, {Langer}, {Yoon}, {Sanyal},
  {de Mink}, {Evans}, \& {Dermine}}]{Szecsi2015}
{Sz{\'e}csi}, D., {Langer}, N., {Yoon}, S.-C., {et~al.} 2015, \aap, 581, A15

\bibitem[{{Tkachenko}(2015)}]{Tkachenko2015}
{Tkachenko}, A. 2015, \aap, 581, A129

\bibitem[{{Tramper} {et~al.}(2011){Tramper}, {Sana}, {de Koter}, \&
  {Kaper}}]{Tramper2011}
{Tramper}, F., {Sana}, H., {de Koter}, A., \& {Kaper}, L. 2011, \apjl, 741, L8

\bibitem[{{Tramper} {et~al.}(2014){Tramper}, {Sana}, {de Koter}, {Kaper}, \&
  {Ram{\'\i}rez-Agudelo}}]{Tramper2014}
{Tramper}, F., {Sana}, H., {de Koter}, A., {Kaper}, L., \&
  {Ram{\'\i}rez-Agudelo}, O.~H. 2014, \aap, 572, A36

\bibitem[{{Vernet} {et~al.}(2011){Vernet}, {Dekker}, {D'Odorico}, {Kaper},
  {Kjaergaard}, {Hammer}, {Randich}, {Zerbi}, {Groot}, {Hjorth}, {Guinouard},
  {Navarro}, {Adolfse}, {Albers}, {Amans}, {Andersen}, {Andersen}, {Binetruy},
  {Bristow}, {Castillo}, {Chemla}, {Christensen}, {Conconi}, {Conzelmann},
  {Dam}, {de Caprio}, {de Ugarte Postigo}, {Delabre}, {di Marcantonio},
  {Downing}, {Elswijk}, {Finger}, {Fischer}, {Flores}, {Fran{\c{c}}ois},
  {Goldoni}, {Guglielmi}, {Haigron}, {Hanenburg}, {Hendriks}, {Horrobin},
  {Horville}, {Jessen}, {Kerber}, {Kern}, {Kiekebusch}, {Kleszcz}, {Klougart},
  {Kragt}, {Larsen}, {Lizon}, {Lucuix}, {Mainieri}, {Manuputy}, {Martayan},
  {Mason}, {Mazzoleni}, {Michaelsen}, {Modigliani}, {Moehler}, {M{\o}ller},
  {Norup S{\o}rensen}, {N{\o}rregaard}, {P{\'e}roux}, {Patat}, {Pena}, {Pragt},
  {Reinero}, {Rigal}, {Riva}, {Roelfsema}, {Royer}, {Sacco}, {Santin},
  {Schoenmaker}, {Spano}, {Sweers}, {Ter Horst}, {Tintori}, {Tromp}, {van
  Dael}, {van der Vliet}, {Venema}, {Vidali}, {Vinther}, {Vola}, {Winters},
  {Wistisen}, {Wulterkens}, \& {Zacchei}}]{Vernet2011}
{Vernet}, J., {Dekker}, H., {D'Odorico}, S., {et~al.} 2011, \aap, 536, A105

\bibitem[{{Vink} {et~al.}(2001){Vink}, {de Koter}, \& {Lamers}}]{Vink2001}
{Vink}, J.~S., {de Koter}, A., \& {Lamers}, H.~J.~G.~L.~M. 2001, \aap, 369, 574

\bibitem[{{von Zeipel}(1924)}]{vonZeipel1924}
{von Zeipel}, H. 1924, \mnras, 84, 665

\bibitem[{{{\v{S}}koda} {et~al.}(2012){{\v{S}}koda}, {Hadrava}, \&
  {Fuchs}}]{Skoda2012}
{{\v{S}}koda}, P., {Hadrava}, P., \& {Fuchs}, J. 2012, in From Interacting
  Binaries to Exoplanets: Essential Modeling Tools, ed. M.~T. {Richards} \&
  I.~{Hubeny}, Vol. 282, 403--404

\bibitem[{{Wellstein} {et~al.}(2001){Wellstein}, {Langer}, \&
  {Braun}}]{Wellstein2001}
{Wellstein}, S., {Langer}, N., \& {Braun}, H. 2001, \aap, 369, 939

\bibitem[{{Woosley} \& {Heger}(2006)}]{Woosley2006}
{Woosley}, S.~E. \& {Heger}, A. 2006, \apj, 637, 914

\bibitem[{{Yoon} {et~al.}(2006){Yoon}, {Langer}, \& {Norman}}]{Yoon2006}
{Yoon}, S.~C., {Langer}, N., \& {Norman}, C. 2006, \aap, 460, 199

\end{thebibliography}

\appendix

\section{GA result plots} \label{appendix1}

   \begin{sidewaysfigure*}
   \centering
   \includegraphics[width=0.9\linewidth]{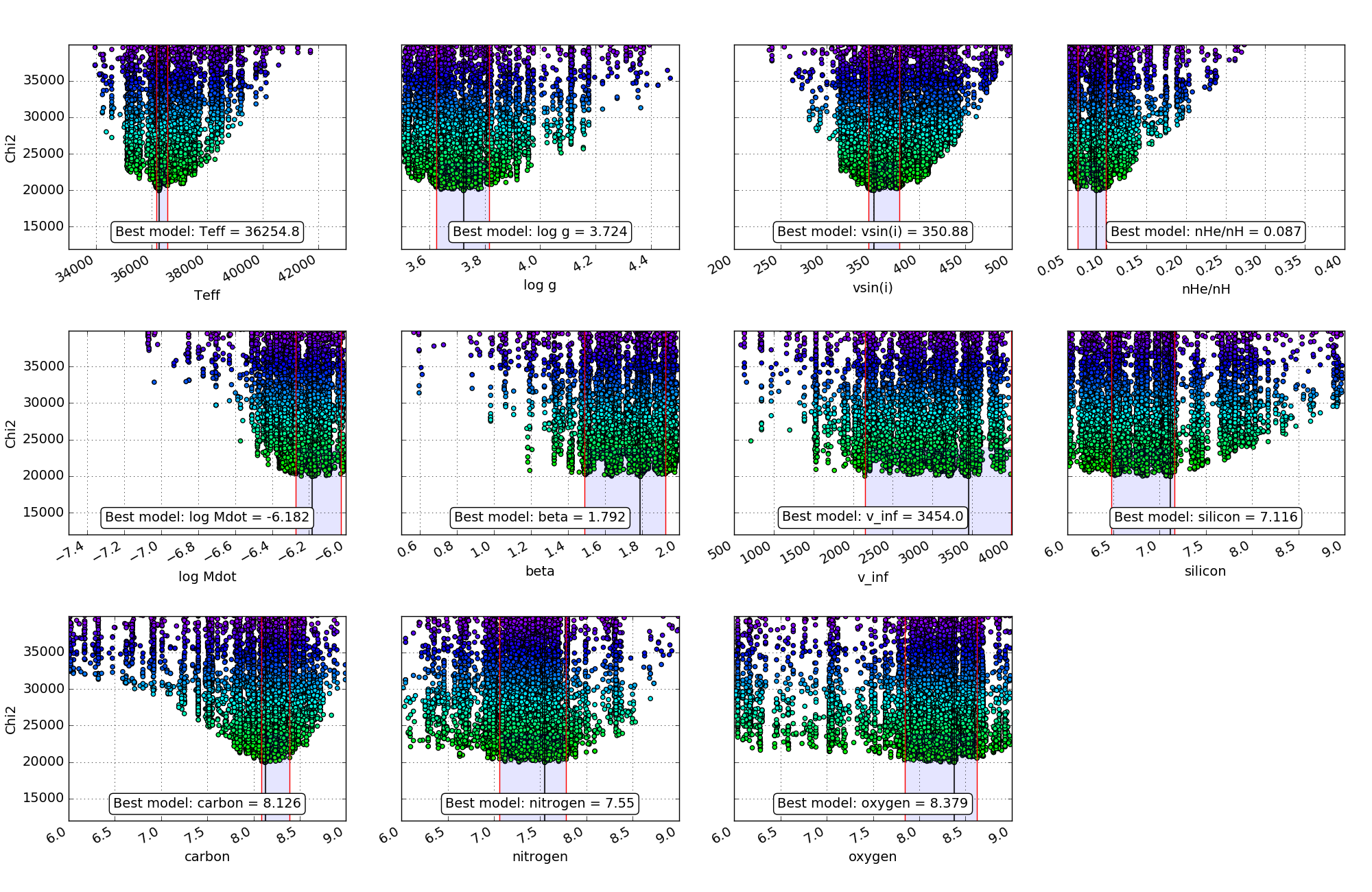}
      \caption{\emph{V382 Cyg Primary component:} 11 dimension $\chi^2$\- merit surface projected along the individual  parameter axis. All computed \textsc{fastwind} models are included.  The confidence interval is indicated with shaded areas while the model with the best $\chi^2$ is indicated by a vertical (black) line within the shaded areas. It is also given on bottom of each panel. 
              }
         \label{ga_fitness_plots-v382_a}
   \end{sidewaysfigure*}

   \begin{sidewaysfigure*}
   \centering
   \includegraphics[width=0.9\linewidth]{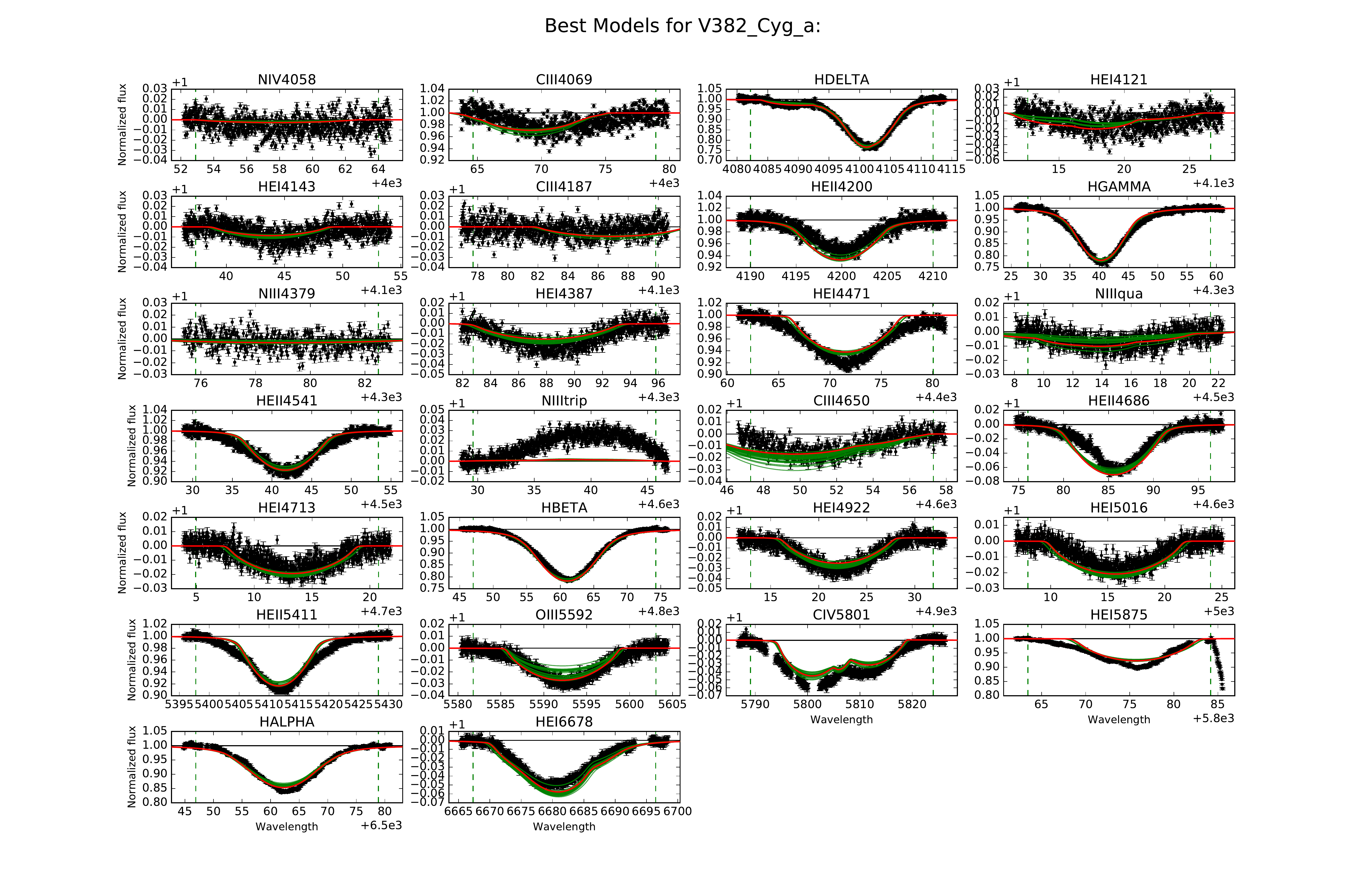}
      \caption{\emph{V382 Cyg Primary component:} line profile fits corresponding to the best fit solution in red and all models in the confidence interval that fulfill the cutoff plotted in green.  The observed spectra and corresponding errors are plotted in black.  The lines are labeled above each panel. 
              }
         \label{ga_lp_plots-v382_a}
   \end{sidewaysfigure*}

   \begin{sidewaysfigure*}
   \centering
   \includegraphics[width=0.9\linewidth]{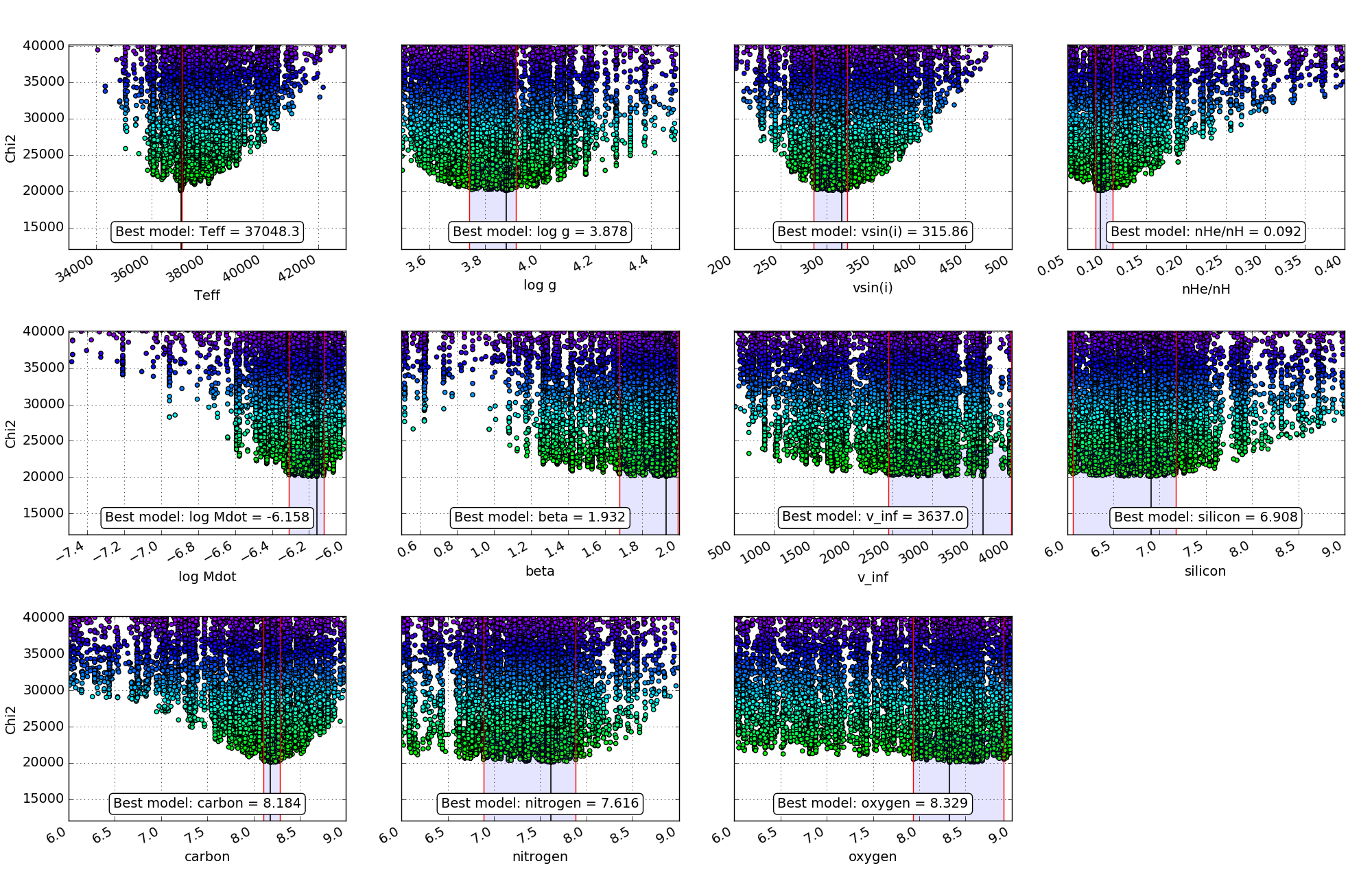}
      \caption{\emph{V382 Cyg Secondary component:} Same as fig. \ref{ga_fitness_plots-v382_a} but for the secondary component of V382 Cyg.
              }
         \label{ga_fitness_plots-v382_b}
   \end{sidewaysfigure*}
   
   \begin{sidewaysfigure*}
   \centering
   \includegraphics[width=0.9\linewidth]{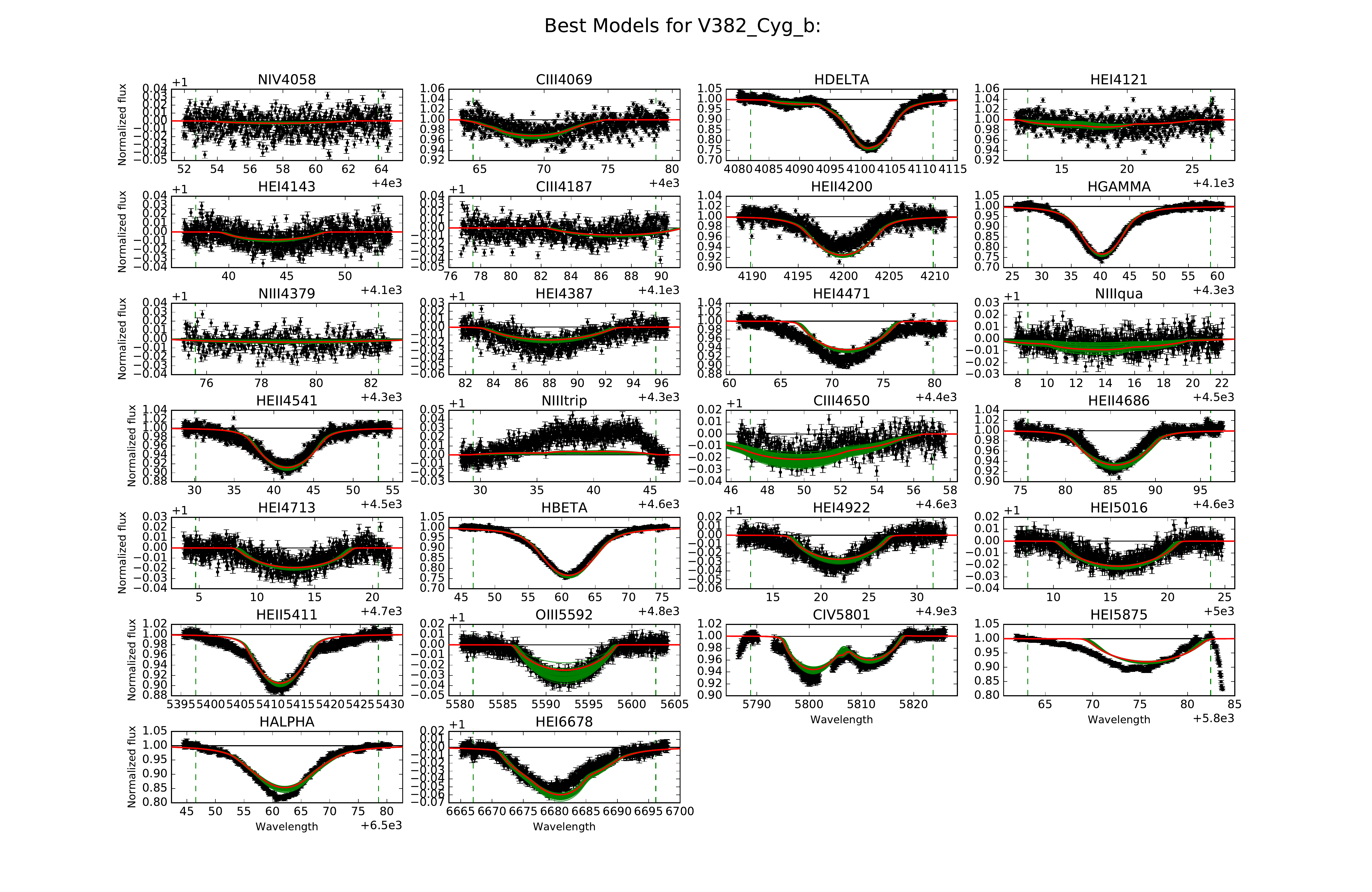}
      \caption{\emph{V382 Cyg Secondary component:} Same as fig. \ref{ga_lp_plots-v382_a} but for the secondary component of V382 Cyg.
              }
         \label{ga_lp_plots-v382_b}
   \end{sidewaysfigure*}
   
   \begin{sidewaysfigure*}
   \centering
   \includegraphics[width=0.9\linewidth]{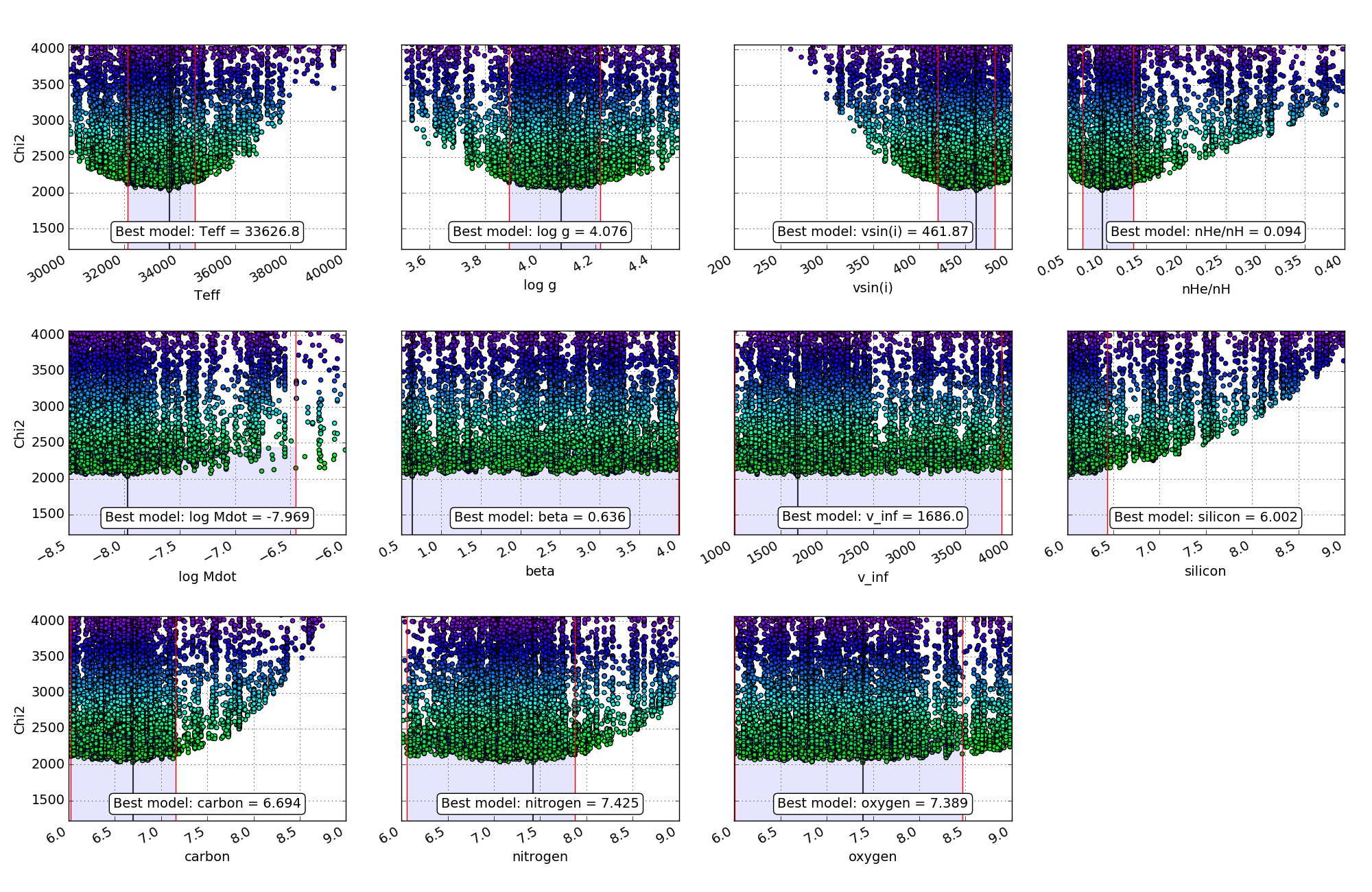}
      \caption{\emph{SMC 108086 Primary component:} Same as fig. \ref{ga_fitness_plots-v382_a} but for the primary component of SMC 108086.
              }
         \label{ga_fitness_plots-smc_a}
   \end{sidewaysfigure*}
   
   \begin{sidewaysfigure*}
   \centering
   \includegraphics[width=0.9\linewidth]{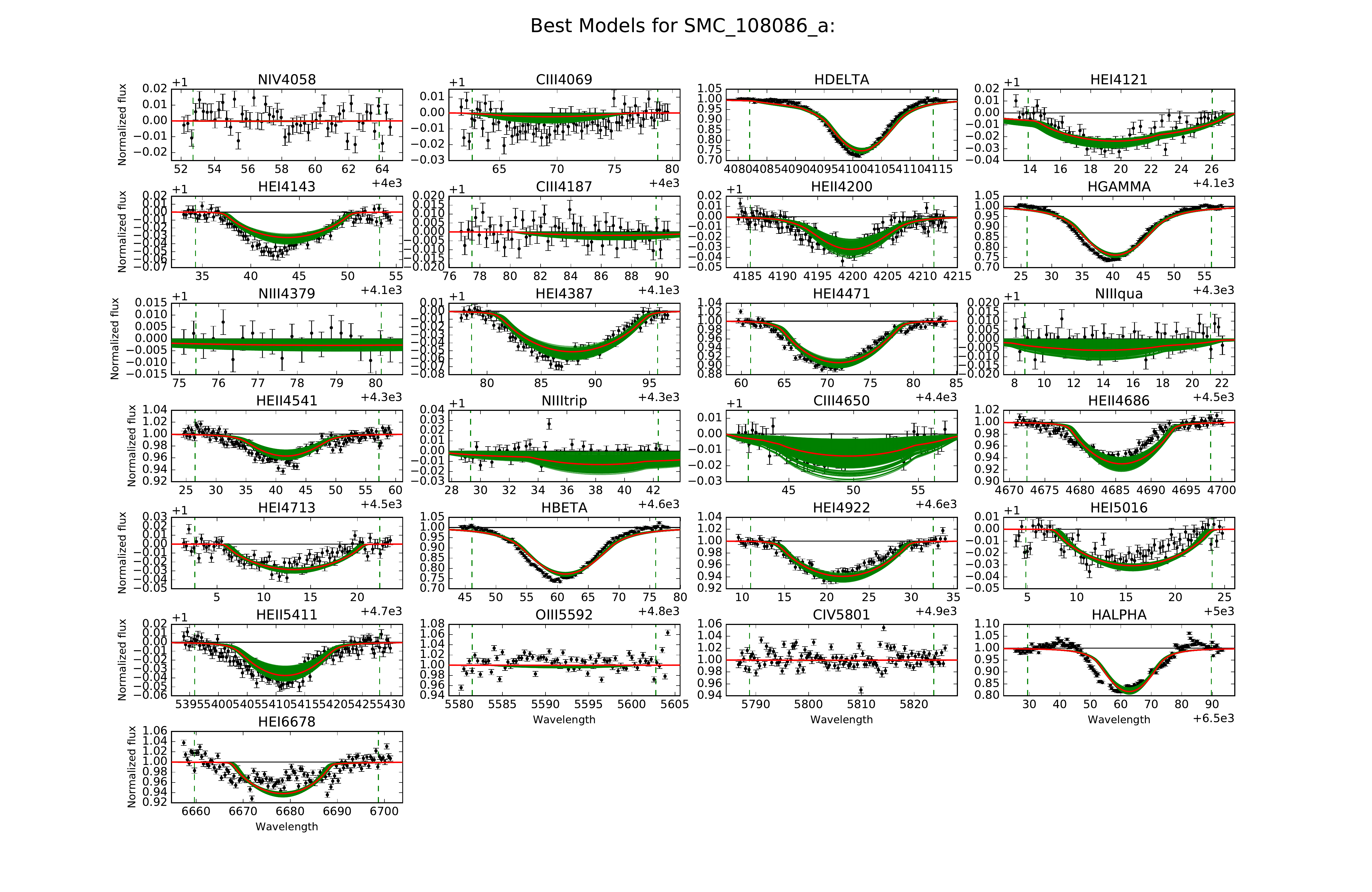}
      \caption{\emph{SMC 108086 Primary component:} Same as fig. \ref{ga_lp_plots-v382_a} but for the primary component of SMC 108086.
              }
         \label{ga_lp_plots-smc_a}
   \end{sidewaysfigure*}
   
   \begin{sidewaysfigure*}
   \centering
   \includegraphics[width=0.9\linewidth]{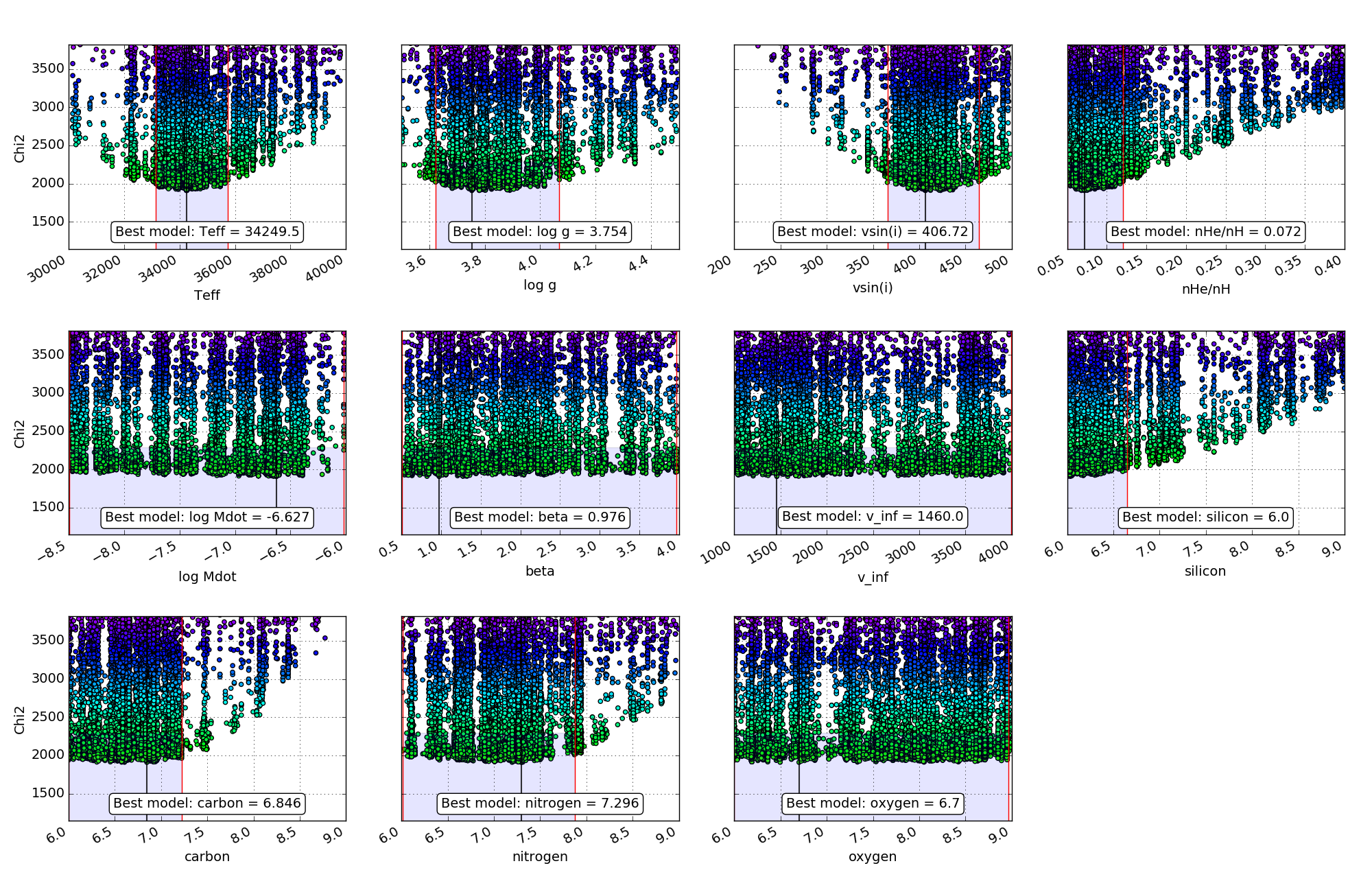}
      \caption{\emph{SMC 108086 Secondary component:} Same as fig. \ref{ga_fitness_plots-v382_a} but for the secondary component of SMC 108086.
              }
         \label{ga_fitness_plots-smc_b}
   \end{sidewaysfigure*}
   
   \begin{sidewaysfigure*}
   \centering
   \includegraphics[width=0.9\linewidth]{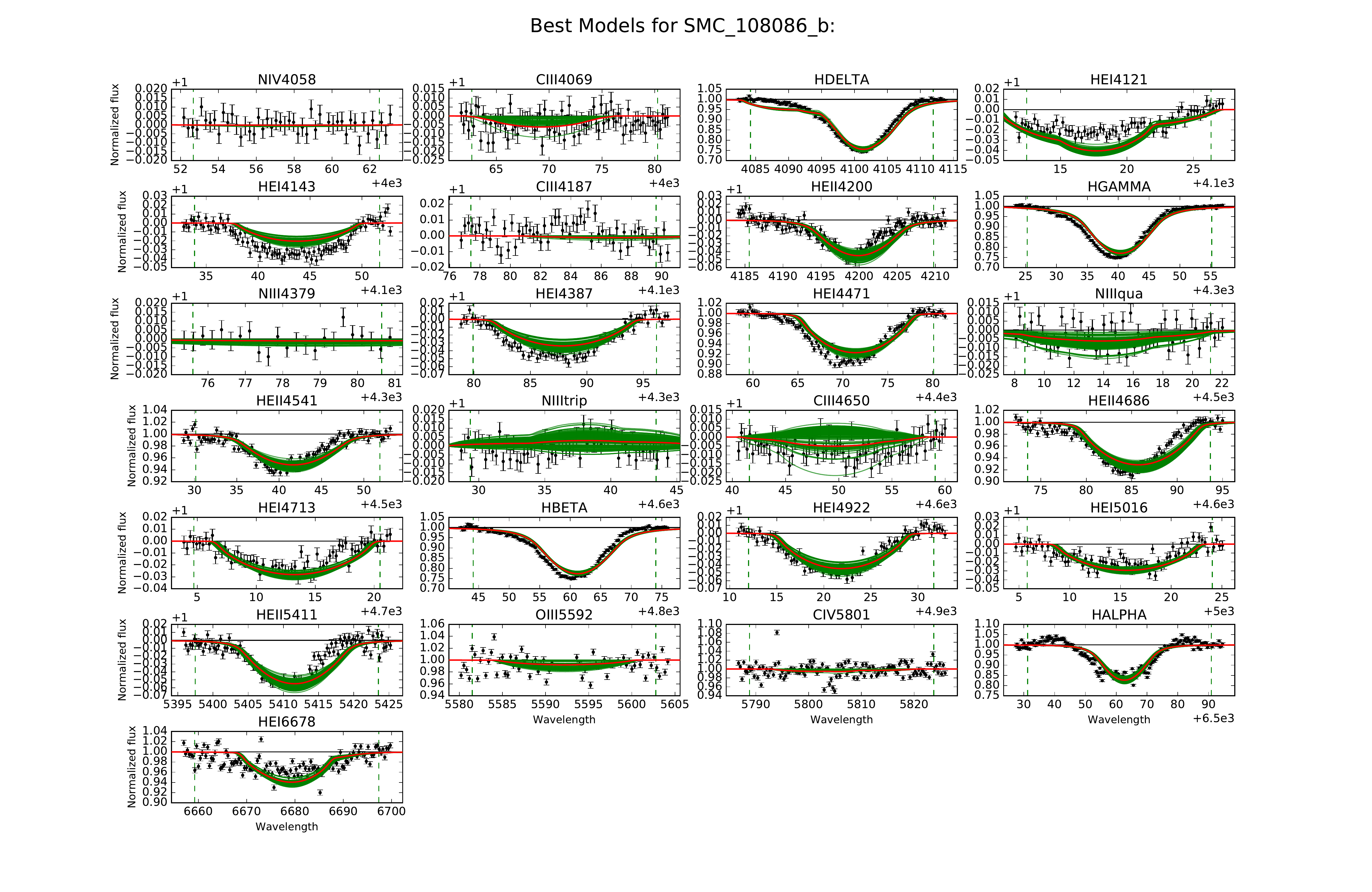}
      \caption{\emph{SMC 108086 Secondary component:} Same as fig. \ref{ga_lp_plots-v382_a} but for the secondary component of SMC 108086.
              }
         \label{ga_lp_plots-smc_b}
   \end{sidewaysfigure*}
   
\section{\spamms result plots} \label{appendix2}

\begin{sidewaysfigure*}
   \centering
   \includegraphics[width=0.9\linewidth]{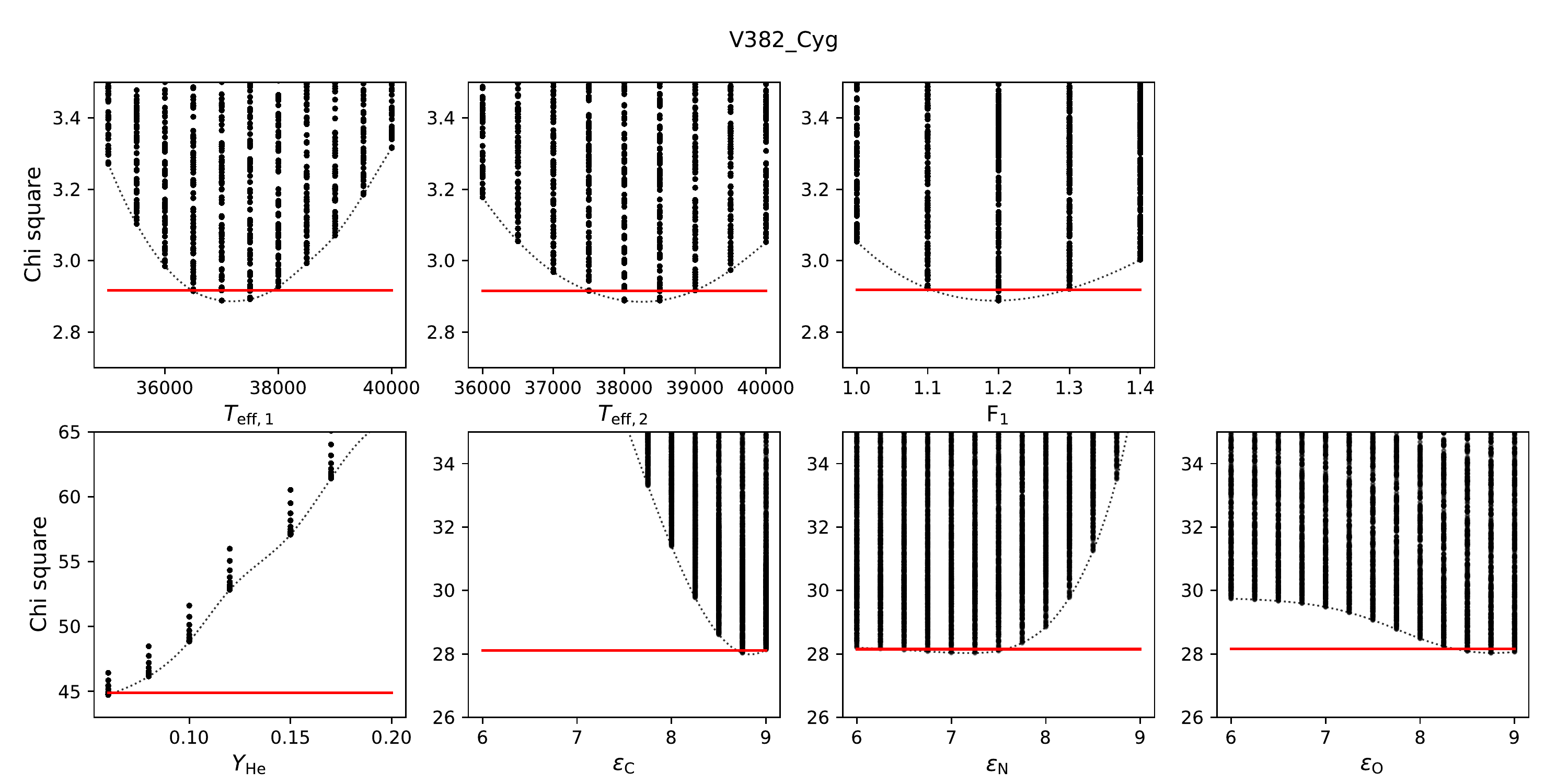}
      \caption{7 dimension $\chi^2$\- merit surface projected along the individual parameter axis for V382 Cyg.  All computed \spamms models are included. This run consists of a combination of three separate optimizations: the first for the temperature and asynchronisity parameter (top row), the second for the helium abundance (bottom row, left most panel) and third for the CNO abundances (bottom row, three right most panels).  The dashed lines trace the $\chi^2$\- merit surface and the red horizontal line indicates the 1-sigma limit.
              }
         \label{spamms_chi2_plots-v382}
   \end{sidewaysfigure*}

 \begin{sidewaysfigure*}
   \centering
   \includegraphics[width=0.9\linewidth]{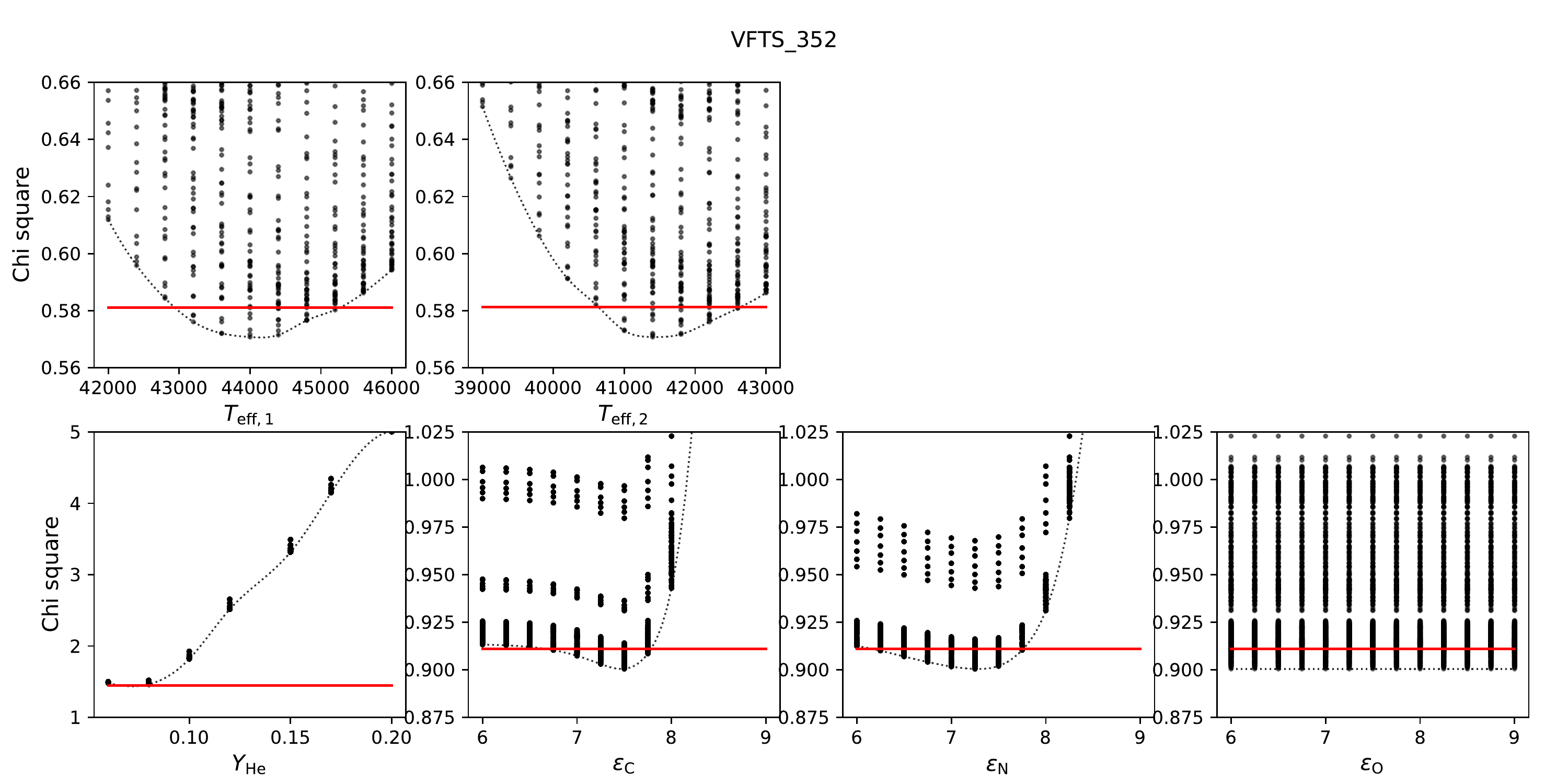}
      \caption{Same as Fig. \ref{spamms_chi2_plots-v382}, but for VFTS 352.  In this case, the asynchronisity parameters were not included in the optimization.
              }
         \label{spamms_chi2_plots-vfts}
   \end{sidewaysfigure*}

 \begin{sidewaysfigure*}
   \centering
   \includegraphics[width=0.9\linewidth]{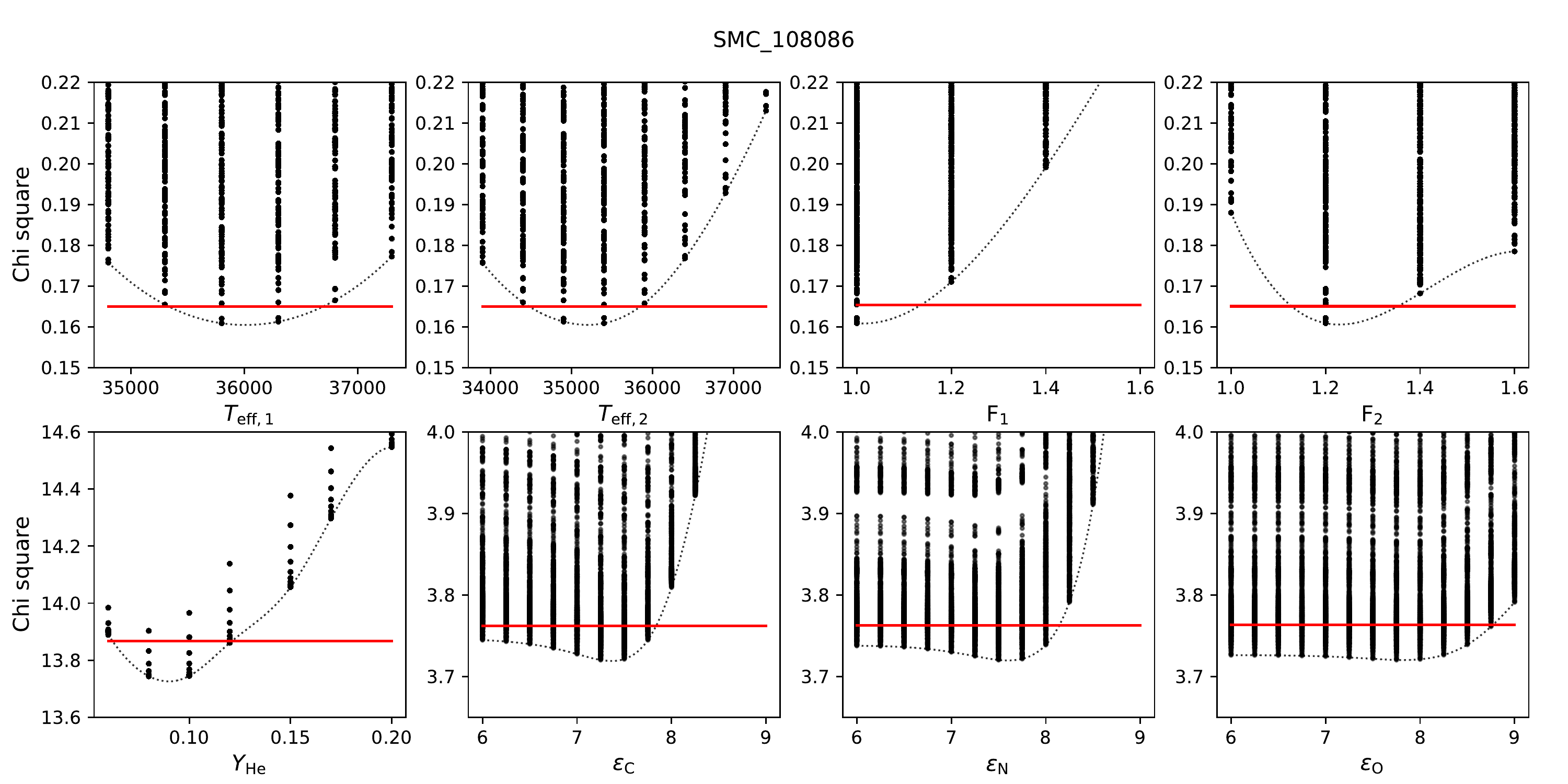}
      \caption{Same as Fig. \ref{spamms_chi2_plots-v382}, but for SMC 108086.  In this case, the asynchronisity parameters were included in the optimization for both the primary and secondary.
              }
         \label{spamms_chi2_plots-smc}
   \end{sidewaysfigure*}

\end{document}